\documentclass[preprint,showpacs,preprintnumbers,amsmath,amssymb,superscriptaddress, nofootinbib]{revtex4}  
\usepackage{graphicx,color}
\usepackage{amsmath,amssymb}
\usepackage{url}
\usepackage{epstopdf}
\usepackage{slashed}

\newcommand{\be}{\begin{equation}}
\newcommand{\ee}{\end{equation}}
\newcommand{\bea}{\begin{eqnarray}}
\newcommand{\eea}{\end{eqnarray}}
\newcommand{\nn}{\nonumber}
\newcommand{\crn}{\nonumber \\}

\newcommand{\bc}{\begin{center}}
	\newcommand{\ec}{\end{center}}

\newcommand {\ba}{\begin{array}}
	\newcommand {\ea}{\end{array}}
\newcommand{\ben}{\begin{enumerate}}
	\newcommand{\een}{\end{enumerate}}
	
\usepackage{epsfig,graphicx} 
\usepackage{bm}
\usepackage{dcolumn}
\begin{document}

\title{Decays $Z\to e_ae_b$ in a 3-3-1 model with neutral leptons}

%

\author{T.T. Hong}\email{tthong@agu.edu.vn}
\affiliation{An Giang University, Long Xuyen City, Vietnam} 
\affiliation{Vietnam National University, Ho Chi Minh City, Vietnam} 
\author{L. T.  Hue }\email{lethohue@vlu.edu.vn}
\affiliation{Subatomic Physics Research Group, Science and Technology Advanced Institute, Van Lang University, Ho Chi Minh City, Vietnam}
\affiliation{Faculty of Applied Technology, School of Technology,  Van Lang University, Ho Chi Minh City, Vietnam} 

\author{L.T.T. Phuong }\email{lttphuong@agu.edu.vn}

\affiliation{An Giang University, Long Xuyen City, Vietnam} 
\affiliation{Vietnam National University, Ho Chi Minh City, Vietnam} 

\author{N.H.T. Nha }
\email{nguyenhuathanhnha@vlu.edu.vn}
\affiliation{Subatomic Physics Research Group, Science and Technology Advanced Institute, Van Lang University, Ho Chi Minh City, Vietnam}
\affiliation{Faculty of Applied Technology, School of Technology,  Van Lang University, Ho Chi Minh City, Vietnam} 

\author{T. Phong Nguyen \footnote{Corresponding author}}\email{thanhphong@ctu.edu.vn}
\affiliation{Department of Physics, Can Tho University, 	3/2 Street, Can Tho, Vietnam}

\begin{abstract}%
We investigate the 3-3-1 model with neutral leptons, a specific extension of the 3-3-1 models {that includes} heavy neutral leptons, {which are sufficient} to accommodate the neutrino oscillation data {via} the inverse seesaw mechanism. We will point out that this model can simultaneously explain the lepton flavor violating decays of the $Z$ and Standard Model-like Higgs  bosons into different charged leptons $Z,h \to e_a e_b$, {as well as} the charged lepton decays $e_b\to e_a \gamma$, in agreement with the recent experimental data. In addition, the numerical results show strong correlations among the decay rates of {the} $Z$ and $h$ {bosons} predicted by this model. {Specifically}, some of {these rates} are nearly linearly dependent on each other. {Consequently}, the decay channels can be {theoretically determined} if one of them is detected in experiments. 
\end{abstract}


\maketitle
\section{\label{intro} Introduction}
\allowdisplaybreaks
We will use  the results of  the general one-loop contributions introduced recently \cite{Hong:2024yhk} to the lepton flavor violating (LFV) decay amplitudes of the $Z$ boson  (LFV$Z$),  $Z \to e_a e_b$, with $e_a, e_b = e, \mu, \tau$ and $e_a\neq e_b$,  to study an extension of  the 3-3-1 {model} adding neutral leptons as $SU(3)_L$ singlets  (331$NL$)  to accommodate the neutrino oscillation data through the inverse seesaw (ISS) mechanism  \cite{Boucenna:2015zwa}. The 3-3-1 models were constructed based on the gauge group $SU(3)_C\otimes SU(3)_L \otimes U(1)_X$ \cite{Singer:1980sw, Pleitez:1992xh,Ozer:1995xi, Foot:1994ym, Frampton:1992wt}, {providing} an interesting explanation {for} the existence three fermion families based on the {anomaly-free} requirements  in the fermion sector \cite{Frampton:1992wt}. Normally, all left-handed fermions in the SM doublets are included in the respective $SU(3)_L$  triplets or antitriplets  of the 3-3-1 models. The  leptons with certain electric charges  in the third component of the lepton $SU(3)_L$ (anti) triplets  are used to distinguish different 3-3-1 models. The 3-3-1 models with {arbitrary electric charges} for these new leptons were also introduced \cite{Diaz:2003dk,Diaz:2004fs,Buras:2012dp}. The 331$NL$ {model discussed} here consists of three neutral left-handed leptons in the third component of the three lepton triplets. Therefore, three other neutral lepton $SU(3)_L$ singlets are {sufficient} for generating active neutrino masses and mixing parameters through the ISS mechanism. It {has been}  shown that if all heavy ISS neutrino masses are degenerate, the strong correlations between {the} three cLFV decay rates (branching ratios)  Br$(e_b\to e_a \gamma)$ will predict invisible signals of the two decays $\tau \to \gamma e,  \gamma \mu$ in the near future \cite{Boucenna:2015zwa}. In contrast, the model with non-degenerate masses of heavy neutrinos will  predict large decay rates  {for} the cLFV and LFV decays of the SM-like Higgs boson (LFV$h$)   $h \to e \tau, \mu \tau$,  which  can reach values of $\mathcal{O}(10^{-4})$ \cite{Hong:2022xjg}.  In this work, we consider the more general situation  of  heavy ISS neutrinos having non-degenerate masses, so that all cLFV decay rates will  reach recent and future experimental sensitivities. More interestingly, two  processes,  LFV$Z$ and LFV$h$ decays, will be discussed simultaneously to guarantee whether their decay rates are excluded, invisible,  or still promising for {search in} upcoming experiments,  
see a summary in Table~\ref{BRdecay} .  
 \begin{table}[ht]  \footnotesize{
 		\begin{tabular}{|c|c|c|c|}
 			\hline
 			& Br & Latest experiment & Future sensitivity\\
 			\hline
 			& $\mathrm{Br}(\mu\rightarrow e\gamma)$ & $< 4.2\times 10^{-13}$ \cite{MEG:2016leq, Belle:2021ysv,  MEGII:2023ltw} & $<6\times 10^{-14}$ \cite{MEGII:2018kmf, Belle-II:2018jsg} \\
 			cLFV & $\mathrm{Br}(\tau\rightarrow \mu\gamma)$ & $<4.4\times 10^{-8}$ \cite{MEG:2016leq, Belle:2021ysv, BaBar:2009hkt} & $< 6.9 \times 10^{-9}$ \cite{MEGII:2018kmf, Belle-II:2018jsg} \\
 			& $\mathrm{Br}(\tau\rightarrow e\gamma)$ & $<3.3\times 10^{-8}$ \cite{MEG:2016leq, Belle:2021ysv, BaBar:2009hkt} & $< 9.0 \times 10^{-9}$ \cite{MEGII:2018kmf, Belle-II:2018jsg} \\
 			\hline
 			& $\mathrm{Br}(h\rightarrow \mu^\pm e^\mp)$ & $<6.1\times 10^{-5} $\cite{ATLAS:2019old} &  $\sim\mathcal{O}(10^{-5}) $ \cite{Qin:2017aju} \\
 			LFV$h$ & $\mathrm{Br}(h\rightarrow \tau^\pm \mu^\mp)$ & $<1.5\times 10^{-3} $\cite{CMS:2021rsq} &  $\sim\mathcal{O}(10^{-4}) $ \cite{Qin:2017aju, Barman:2022iwj, Aoki:2023wfb} \\
 			& $\mathrm{Br}(h\rightarrow \tau^\pm e^\mp)$ & $<2.2\times 10^{-3}$\cite{CMS:2021rsq} &  $\sim\mathcal{O}(10^{-4}) $ \cite{Qin:2017aju} \\
 			\hline
 			& $\mathrm{Br}(Z\rightarrow \mu^\pm e^\mp)$ & $<2.62\times 10^{-7} $\cite{ATLAS:2022uhq} & $\sim 7\times 10^{-8}$  (HL-LHC) and $10^{-10}$ (FCC-ee) \cite{ATLAS:2021bdj, Dam:2018rfz, FCC:2018byv} \\
 			LFV$Z$ & $\mathrm{Br}(Z\rightarrow \tau^\pm \mu^\mp)$ & $<6.5\times 10^{-6} $\cite{ATLAS:2021bdj} & $\sim 10^{-6}$  (HL-LHC)  and $10^{-9}$ (FCC-ee)  \cite{Dam:2018rfz, FCC:2018byv} \\
 			& $\mathrm{Br}(Z\rightarrow \tau^\pm e^\mp)$ & $<5.0\times 10^{-6}$\cite{ATLAS:2021bdj} & $\sim 10^{-6}$  (HL-LHC)  and $10^{-9}$ (FCC-ee) \cite{Dam:2018rfz, FCC:2018byv} \\
 			\hline
 		\end{tabular}
 		\caption{ The latest and expected experimental sensitivities on the LFV branching ratios (Br).    \label{BRdecay}} } 
 \end{table}
 
LFV$Z$ decays have been investigated in many Beyond the Standard Models (BSM), predicting  large LFV$Z$ but small LFV$h$ decay rates compared to upcoming experimental sensitivities. For example, it was shown that  Br$(Z \to  \mu^\mp \tau^\pm) \varpropto \mathcal{O}(10^{-5}), $ and Br$(h \to  \mu^\mp e^\pm) \varpropto \mathcal{O}(10^{-11})$ in a simple SM extension adding only ISS neutrinos  \cite{Abada:2022asx}. Similarly, a recent discussion on a Two-Higgs Doublet Model  (2HDM) with ISS neutrinos predicts the large Br$(Z \to \mu^\pm \tau^\mp ) \varpropto \mathcal{O}(10^{-8})$  
	but suppressed  values of Br$(h \to \mu^\pm e^\mp) < 10^{-8}$ \cite{Hong:2023rhg} in the regions of parameter space accommodating the $(g-2)_{e,\mu}$ data.  When ignoring the $(g-2)_{e_a}$ anomalies, another investigation on a 2HDM with standard seesaw neutrinos predicts promising signals of LFV$h$, but the LFV$Z$ decay rates are suppressed to be observed experimentally \cite{Jurciukonis:2021izn}. This implies that combined results from experimental searches for these two LFV$h$ and $Z$ decays may provide useful information to determine the reality of available BSM models. Therefore, LFV$Z$ decays are attractive objects {for both} theoretical studies   \cite{Hong:2024yhk, Hong:2023rhg, Jurciukonis:2021izn, Grimus:2002ux,Herrero:2018luu}  and experimental searches  \cite{DELPHI:1996iox, OPAL:1995grn,  ATLAS:2014vur}. In conclusion, {studying} LFV$Z$ decays is necessary for the 331$NL$ model to see whether they are the strictest constraint among the three classes of LFV decay rates we focus on in this work.

The structure of this work is organized as follows. In section \ref{model}, we will review the necessary components of the 331$NL$ model for studying the $Z \to e_a e_b$ decays and how the ISS framework can produce the active neutrino masses and mixing  parameter{s} consistent with the experimental data.  We will also discuss the one-loop contributions to the decay amplitudes  $Z\to e_b e_a$. The couplings and analytic formulas related to the LFV$Z$ decays will be determined in section \ref{sec_coupling}. In section IV, we will {numerically investigate} three classes of cLFV, LFV$Z$, and LFV$h$ decays, working out the allowed regions of the parameter space satisfying all LFV experimental limits. In section V, crucial results will be summarized and presented in our conclusions. Finally, Appendix A shows the master functions of the one-loop contribution to LFV$Z$ decay amplitudes in the unitary gauge.

\section{\label{model} The 3-3-1 model with neutral lepton}
\subsection{Particle content and neutrino masses from the ISS mechanism}
The 331$NL$ model is a specific version  of  the 3-3-1 models, {featuring} new neutral leptons in the third component of the  left-handed lepton triplets. We {focus solely} on particle spectrum of the 331$NL$ model {relevant} to this work.  
In the 331$NL$ model, {the electric charge operator} $Q=T_3-\frac{T_8}{\sqrt{3}}+X$ is identified by the gauge group  $SU(3)_L\times U(1)_X$, in which $T_{3,8}$ are diagonal $SU(3)_L$ generators. Each lepton generation contains a $SU(3)_L$ triplet $L_{a}= (\nu_a,~e_a, N_a)_L^T\sim (3,-\frac{1}{3})$, {a RH (right-handed) charged lepton $e_{aR}\sim (1,-1)$}, and a new neutral lepton $X_{aR} \sim (1,0)$  with $a=1,2,3$. The three Higgs triplets of the model are    $\rho=(\rho^+_1,~\rho^0,~\rho^+_2)^T\sim (3,\frac{2}{3})$,
$\eta=(\eta_1^0,~\eta^-,\eta^0_2)^T\sim (3,-\frac{1}{3})$, and $\chi=(\chi_1^0,~\chi^-,\chi^0_2)^T\sim (3,-\frac{1}{3})$.  All quark and lepton masses at tree-level are generated by the vacuum expectation values (vev) are  $\langle\rho \rangle=(0,\,\frac{v_1}{\sqrt{2}},\,0)^T$, $\langle \eta \rangle=(\frac{v_2}{\sqrt{2}},\,0,\,0)^T$ and $\langle \chi \rangle=(0,\,0,\,\frac{w}{\sqrt{2}})^T$. Two neutral Higgs components have zero vevs due to they have non-zero  lepton numbers~\cite{Hue:2021xap}, when the 331$NL$ model respects a global symmetry $U(1)_\mathsf{L}$ defined as the  generalized lepton number $\mathsf{L}$ that relates to the normal lepton number by  $L=\frac{4}{\sqrt{3}}T^8 +\mathsf{L}$ \cite{Chang:2006aa}. Here, we explain more precisely that the two triplets $\eta$ and $\chi$ have the same quantum numbers of the total gauge group, but their  values of the $\mathsf{L}$ assignments are different. Therefore, they have different Yukawa couplings and {vevs}, as indicated in   Ref. \cite{Chang:2006aa}.


The covariant kinetic Lagrangian of the Higgs boson triplets generates mass for nine gauge bosons $\mathcal{L}^{\phi}=\sum_{\phi=\chi,\eta,\rho} \left(D_{\mu}\phi\right)^{\dagger}\left(D^{\mu}\phi\right)$, where 
$D_\mu  = \partial _\mu  - i g  {W}_\mu ^a{T^a} - i{g_X}{T^9}X {X_\mu }$, {with} $a=1,2,..,8$, and    $T^9 \equiv \frac{I_3}{\sqrt{6}}$ and $\frac{1}{\sqrt{6}}$ corresponding to (anti)triplets and singlets \cite{Buras:2012dp}. 
The matrix $W^aT^a$ can be presented as
\bea W^a_{\mu}T^a=\frac{1}{2}\left(
\begin{array}{ccc}
	W^3_{\mu}+\frac{1}{\sqrt{3}} W^8_{\mu}& \sqrt{2}W^+_{\mu} &  0 \\
	\sqrt{2}W^-_{\mu} &  -W^3_{\mu}+\frac{1}{\sqrt{3}} W^8_{\mu} & \sqrt{2}Y^{+}_{\mu} \\
	0& \sqrt{2}Y^{-}_{\mu} &-\frac{2}{\sqrt{3}} W^8_{\mu}\\
\end{array}
\right),
\label{wata1}\eea 
where $T^a =\lambda_a/2$ {is} respective to a triplet representation. The model consists of $W^{\pm}$  and $Y^{\pm}$, which are two pairs of singly charged gauge bosons, as follows
\begin{align}
W^{\pm}_{\mu}&=\frac{1}{\sqrt{2}}(W^1_{\mu}\mp i W^2_{\mu}),\; Y^{\pm}_{\mu}=\frac{1}{\sqrt{2}}(W^6_{\mu}\pm i W^7_{\mu}),\; \;  	
\label{singlyG}
\end{align}
with the respective  masses  $m_W^2=\frac{g^2}{4}\left(v_1^2+v_2^2\right)$ and $	m_Y^2=\frac{g^2}{4}\left(w^2+v_1^2\right)$. The model will be broken step by step as $SU(3)_L\times U(1)_X \xrightarrow{w} SU(2)_L\times U(1)_Y \xrightarrow{v_1, v_2} U(1)_Q$, {where} the vevs satisfy the hierarchy $w\gg v_1,v_2$. {This leads to identifying}  $W^{\pm}$ as  the SM ones. Therefore the following relations are obtained   as follows \cite{Chang:2006aa}
\begin{align} \label{eq_SMi}
	v_2^2+v_1^2\equiv v^2=(246 \mathrm{GeV})^2, \quad  \frac{g_X}{g}= \frac{3\sqrt{2}s_W}{\sqrt{3-4s^2_W}},\quad  s_W=\frac{e}{g},
\end{align}
where $e$ and $s_W$ are the electric charge and sine of the Weinberg angle $s^2_W\simeq 0.231$ \cite{ParticleDataGroup:2022pth}, respectively. Similarly to the 2HDM,  the following parameters will be used
\begin{equation}\label{eq_tbeta}
	t_{\beta}\equiv \tan\beta=\frac{v_2}{v_1}, \quad v_1=c_{\beta}\times v, \quad v_2=s_{\beta} \times v.
\end{equation}
The lepton masses generated from the Yukawa Lagrangian are below 
\begin{align}
	\label{eq_Lylepton}
	\mathcal{L}^{\mathrm{Y}}_l =&-h^e_{ab}\overline{L_{a}}\rho e_{bR}+
	h^{\nu}_{ab} \epsilon^{ijk} \overline{(L_{a})_i}(L_{b})^c_j\rho^*_k 	
	%
	- y^{\chi}_{ba}\overline{ X_{bR}}\chi^{\dagger} L_{a} -\frac{1}{2} (\mu_{X})_{ab}\overline{X_{aR}} \left(X_{bR}\right)^c
+\mathrm{h.c.},
\end{align}
where $a,b=1,2,3$. 
 The charged lepton masses generated from the first term of Eq. \eqref{eq_Lylepton} as $	m_{e_a}\equiv \frac{h^e_{ab}v_1}{\sqrt{2}} \delta_{ab}$, {where we assume that} the flavor states are physical. The 331$NL$ model under consideration inherits lepton sector and LFV couplings given in Eq. \eqref{eq_Lylepton}, {which} are completely different from the class of the 3-3-1 models with six neutral lepton singlets  mentioned in Ref. \cite{Hong:2024yhk}. Namely, the 331$NL$ consists of  three exotic neutrinos in the  $SU(3)_L$ lepton triplets. Three other $SU(3)_L$ singlets of neutral leptons are needed to work the ISS mechanism.

In the basis $n'_{L}=(\nu_{L}, N_{L}, (X_R)^c)^T$,   Eq. \eqref{eq_Lylepton} gives a  neutrino mass term with  the total $9\times 9$ mass matrix written in terms of  the following  block form of $3\times 3$ sub-matrices  \cite{Nguyen:2018rlb} 
\begin{align}
	-{\mathcal{L}}^{\nu}_{\mathrm{mass}}=\frac{1}{2}\overline{(n'_L)^c}\mathcal{M}^{\nu}n'_L  +\mathrm{h.c.}, \,\mathrm{ where }\quad \mathcal{M}^{\nu}=\begin{pmatrix}
		\mathcal{O}_{3}	& m^T_{D} & 	\mathcal{O}_{3}\\
		m_{D}	&\mathcal{O}_{3}	  & M^T_R \\
		\mathcal{O}_{3}& M_R& \mu_X
	\end{pmatrix},  \label{Lnu1}
\end{align}
where $(n'_{L})^c=((\nu_L)^c, (N_L)^c, X_R)^T$,  $(M_R)_{ab}\equiv y^{\chi}_{ab}\frac{w}{\sqrt{2}}$, and $(m^T_D)_{ab}= -(m_D)_{ab}\equiv \sqrt{2}v_1h^{\nu}_{ab}$  with $a,b=1,2,3$. The matrix  $\mu_{X}$  in Eq.~\eqref{eq_Lylepton}  is symmetric, so we consider it a diagonal one, simplifying the calculations without loss of generality.

By using the $9\times9$ unitary matrix $U^{\nu}$ that can diagonalize mass matrix $\mathcal{M}^{\nu}$ as follows
\begin{align}
	U^{\nu T}\mathcal{M}^{\nu}U^{\nu}=\hat{M}^{\nu}=\mathrm{diag}(m_{n_1},m_{n_2},..., m_{n_{9}})=\mathrm{diag}(\hat{m}_{\nu}, \hat{M}_N), \label{diaMnu}
\end{align}
where the physical states $n_{iL}$ have masses corresponding to $m_{n_i}$ ($i=1,2,\dots 9$). The two mass matrices    $\hat{m}_{\nu}=\mathrm{diag}(m_{n_1},\;m_{n_2},\;m_{n_3})$  and   $\hat{M}_N$ $=\mathrm{diag}(m_{n_4},\;m_{n_5},...,\;m_{n_{9}})$, {correspond} to  the  active extra neutrinos  $n_{aL}$ ($a=1,2,3$) and  $n_{IL}$ ($I=1,2,..,6$), respectively.  
The following approximate solution for the neutrino mixing matrix $U^{\nu}$ is reasonable for any particular seesaw mechanisms:
\begin{align}
	\label{eq_Unu}
	U^{\nu}= \Omega \left(
	\begin{array}{cc}
		U_{\mathrm{PMNS}} & \mathcal{O}_{3\times6}\\
		\mathcal{O}_{6\times3} & V \\
	\end{array}
	\right), \;\; \Omega \simeq 
	\left(
	\begin{array}{cc}
		I_3-\frac{1}{2}RR^{\dagger} & R \\
		-R^\dagger &  I_6-\frac{1}{2}R^{\dagger} R\\
	\end{array}
	\right),
\end{align}
where $U_{PMNS}$ is the Pontecorvo-Maki-Nagakawa-Sakata (PMNS) matrix, and $R$, $V$ are $3\times6$,  $3\times6$ matrices, respectively.   To derive the ISS relations perturbatively, it is assumed that all entries of the matrix $R$ satisfy $|R_{aI}|\ll$1.  

The flavor and mass eigenstates are shown through the {following relation:}
\begin{equation}
	n'_L=U^{\nu} n_L, \quad  \mathrm{and} \; (n'_L)^c=U^{\nu*}  (n_L)^c \equiv U^{\nu*}  n_R, \label{Nutrans}
\end{equation}
where $n_L\equiv(n_{1L},n_{2L},...,n_{9L})^T$, and $n_i$ represents the Majorana states with components $(n_{iL},\;n_{iR})^T$. 

We summarize here the ISS expressions:
\begin{align}
	R^*_2&=m_D^TM^{-1}_R,  \; 
	R^*_1 = -R^*_2 \mu_X\left( M^T_R\right)^{-1}\simeq \mathcal{O}_{3\time3},  \label{eq_R12}
	\\  m_{\nu} &=R_2^*\mu_XR_2^{\dagger} = U^*_{\mathrm{PMNS}} \hat{m}_{\nu}U^\dagger_{\mathrm{PMNS}} =m_D^TM^{-1}_R\mu_X\left( M^{-1}_R\right)^Tm_D,  \label{eq_mnu}
	\\ V^*\hat{M}_NV^{\dagger}&=M_N +\frac{1}{2} M_N R^{\dagger}R +	\frac{1}{2}R^{T}R^*M_N.  \label{eq_MN}
\end{align} 
All independent parameters $x_{12,13}$ and three entries of $M^{-1}\equiv M^{-1}_R\mu_X\left( M^{-1}_R\right)^T$ can be determined from experimental data of $m_{\nu}$ \cite{Boucenna:2015zwa, Nguyen:2018rlb}. {In} particular the Dirac mass matrix exhibits an antisymmetric structure, {given by}
\begin{equation} \label{eq_mD}
	m_D= ze^{i\alpha_{23}}\times  \tilde{m}_D,\; \text{where}~ \tilde{m}_D=\begin{pmatrix}
		0&x_{12}  &x_{13}  \\ 
		-x_{12}& 0 &1  \\ 
		-x_{13}& -1 &0 
	\end{pmatrix}, 
\end{equation}
where $ \alpha_{23}\equiv \arg[h^{\nu}_{32}]$ and 
 \begin{equation}\label{eq_zdef}
	z=\sqrt{2}|h^{\nu}_{32}|v_1 = \sqrt{2}|h^{\nu}_{23}|v_1\,\equiv c_{\beta}z_0	
\end{equation}
is a real and positive parameter.  We note that $z_0=\sqrt{2}v|h^{\nu}_{23}|$ is considered as the scale of the Dirac mass matrix $m_D$. The pertubative limit of the Yukawa coupling $h^{\nu}_{23}$ leads to an upper constraint that  $z_0<1223$ GeV \cite{Nguyen:2018rlb}.  The phase $\alpha_{23}=0$ is chosen after redefining the phase of $L_a$ in the second term of  Eq. \eqref{eq_Lylepton}. The Eq. \eqref{eq_mnu} gives $\left( m_{\nu}\right)_{ij}=\left[m_D^TM^{-1}m_D\right]_{ij}$ for all $i,j=1,2,3$, {resulting in} six independent functions. 
{By} inserting $\left( m_{\nu}\right)_{ij}$  into the three remaining relations with $i=j$, we obtain 
\begin{align}
	\label{eq_xijM}	
	 x_{12}= \frac{\left(m_{\nu })_{12}\right. \left(m_{\nu })_{13}\right.-\left(m_{\nu })_{11}\right. \left(m_{\nu })_{23}\right.}{\left(m_{\nu })_{13}\right. \left(m_{\nu })_{23}\right.-\left(m_{\nu })_{12}\right.
		\left(m_{\nu })_{33}\right.},\;x_{13}&=\frac{(m_{\nu })_{13}^2-\left(m_{\nu })_{11}\right. \left(m_{\nu})_{33}\right.}{\left(m_{\nu })_{13}\right. \left(m_{\nu })_{23}\right.-\left(m_{\nu })_{12}\right. \left(m_{\nu
		})_{33}\right.}, 
\end{align}
and $\mathrm{Det}[m_{\nu}]=0$. We express all parameter of the matrix $\mu_{X}$ as certain but lengthy functions of $(ze^{i\alpha_{32}})$, all entries of  $M_R$ and $m_{\nu}$ by using $M^{-1}=M^{-1}_R\mu_X\left( M^{-1}_R\right)^T$. While experiment data help determine $m_{\nu}$, others are free parameters.

We consider in the limit that $|R_2|\ll1$, based on Eq. \eqref{eq_MN}, we can {approximately determine} the heavy neutrino masses, namely
\begin{align}
	V^*\hat{M}_NV^{\dagger}\simeq M_N.  \label{eq_MNa}
\end{align}
For convenience, we  identify the reduced matrix:
\begin{align}
	\label{eq:kij}
	M_R\equiv ze^{i\alpha_{23}} \tilde{M}_R, \;  \left(\tilde{M}_R\right)_{ij}\equiv k_{ij}, 
\end{align}
so that $R_2^*=-\tilde{m}_D/\tilde{M}_R$.  Since the matrix $M_R$ is always diagonalized by two unitary transformations $V_L$ and $V_R$ \cite{Dreiner:2008tw},
\begin{align}
	V_L^TM_RV_R= z\times \hat{k}=z \times \mathrm{diag}(\hat{k}_1,\;\hat{k}_2,\;\hat{k}_3), \label{eq_MRd}
\end{align}
where all $\hat{k}_{1,2,3}$ are always positive {and} $\hat{k}_a\gg1$, so that all ISS relations are valid. Therefore,   $M_R$ is determined in terms of  free parameters included in $\hat{k}$, $V_{L,R}$. Based on Eq. \eqref{eq_MN}, we can prove that the matrix $V$ can be determined approximately as follows
\begin{align}
	V=\frac{1}{\sqrt{2}} \begin{pmatrix}
		V_R	&  iV_R\\
		V_L & -iV_L
	\end{pmatrix} \to V^TM_NV= z\times \begin{pmatrix}
		\hat{k}	& \mathcal{O}_{3\times3}\\
		\mathcal{O}_{3\times3} & \hat{k}
	\end{pmatrix}.  \label{eq_V0}
\end{align}
Consequently,  for any qualitative estimations, by using the crude approximations that heavy neutrino masses are  $m_{n_{a+3}}=m_{n_{a+6}}\simeq z \hat{k}_a$ with a=1,2,3; $R_1\simeq \mathcal{O}_3$;  and
\begin{align} \label{eq_Unu1}
	U^{\nu}\simeq \begin{pmatrix} 
		\left( I_3 -\frac{1}{2} R_2R^\dagger_2\right) U_{\mathrm{PMNS}}&  \frac{1}{\sqrt{2}}R_2 V_L& \frac{-i}{\sqrt{2}}R_2 V_L \\
		\mathcal{O}_3& \frac{V_R}{\sqrt{2}} &  \frac{iV_R}{\sqrt{2}}\\
		-R^\dagger_2U_{\mathrm{PMNS}}& \left(I_3 -\frac{R^\dagger_2R_2}{2} \right)\frac{V_R}{\sqrt{2}}  
		& \left(I_3 -\frac{R^\dagger_2R_2}{2}\right)\frac{-iV_R}{\sqrt{2}}
	\end{pmatrix}. 
\end{align}
We have verified that the {mentioned approximations yield} numerical results {consistent with the} calculations discussed in Ref. \cite{Hue:2021xap}. Based on Eq. \eqref{eq_xijM}, the Dirac mass matrix $\tilde{m}_D$ will be determined numerically through  the active neutrino matrix $m_{\nu}$ given in Eq. \eqref{eq_mnu},  using the input of $3\sigma$ ranges of neutrino oscillation data. We investigate the parameter space by scanning the free parameters $z_0$, $\hat{k}_{1,2,3}$, {and} $V_{R}$ within reasonable ranges. Using these values, we construct the total neutrino mixing matrix $U^{\nu}$ defined in Eq. \eqref{eq_Unu1}.  Because $V_L$ satisfies: 
\begin{equation}\label{eq_R2VL}
	R_2V_L=\tilde{m}^{\dagger}_DV^*_R\hat{k}^{-1},	\quad R_2R_2^{\dagger}=\tilde{m}^{\dagger}_DV_R\hat{k}^{-2}\tilde{m}_D,
\end{equation}
which  is {explicitly independent of}  $V_L$, all processes {discussed} here are weakly affected  by $V_L$. {Therefore, $V_L$ will be fixed at $V_L = I_3$ for the remainder of this study}.

The Yukawa Lagrangian generating quark masses {was} presented in Ref.~\cite{Chang:2006aa}. In this work, it is important to emphasize that the perturbative limit  $h^u_{33}<\sqrt{4\pi}$,  leading to a lower bound for $v_2$:  $v_2 >\frac{\sqrt{2}m_t}{\sqrt{4\pi}}$. By combining {this} relationship {with} {Eqs.~\eqref{eq_SMi}} and ~\eqref{eq_tbeta}, the lower bound for $t_{\beta}\ge0.3$. From the tau mass, $m_{\tau}=h^3_{33}\times vc_{\beta}\sqrt{2}\to h^3_{33}=m_{\tau}\sqrt{2}/(v c_{\beta})<\sqrt{4\pi}$, {we derive} the upper bound for $t_{\beta}$, resulting in the rather weak upper bound $t_{\beta}=\sqrt{1/c^2_{\beta}-1}\leq 346$. 

\subsection{Higgs bosons}
In this work, the Higgs potential respects the new general lepton number introduced  in Ref.~\cite{Chang:2006aa}, namely
\begin{align}
	\label{eq_Vh}
	V_{higgs}&= \sum_{\phi} \left[ \mu_\phi^2 \phi^{\dagger}\phi +\lambda_\phi \left(\phi^{\dagger}\phi\right)^2 \right]  + \lambda_{12}(\eta^{\dagger}\eta)(\rho^{\dagger}\rho)
	+\lambda_{13}(\eta^{\dagger}\eta)(\chi^{\dagger}\chi)
	+\lambda_{23}(\rho^{\dagger}\rho)(\chi^{\dagger}\chi)  \crn
	& +\tilde{\lambda}_{12} (\eta^{\dagger}\rho)(\rho^{\dagger}\eta) 
	+\tilde{\lambda}_{13} (\eta^{\dagger}\chi)(\chi^{\dagger}\eta)
	+\tilde{\lambda}_{23} (\rho^{\dagger}\chi)(\chi^{\dagger}\rho) +\sqrt{2}f \omega \left(\epsilon_{ijk}\eta^i\rho^j\chi^k +\mathrm{h.c.} \right), 
\end{align}
where $f$ is a dimensionless term, $\phi=\eta,\rho,\chi$.  
In previous research, the minimum conditions of the Higgs potential and the identification of the SM-like Higgs boson were thoroughly examined, as seen in references such as Refs. \cite{Ninh:2005su, Hue:2015fbb}. Here, we summarize the essential results: The 331$NL$ model includes two pairs of singly charged Higgs boson, denoted as $h^{\pm}_{1}, h^{\pm}_{2}$, along with two Goldstone bosons $G^{\pm}_{W},  G^{\pm}_{Y}$ of the singly charged gauge bosons  $W^{\pm}$, and $Y^{\pm}$, respectively. {The mass terms are}  $m^2_{h^{\pm}_1}=  \left( \frac{\tilde{\lambda}_{12}v^2}{2} +\frac{fw^2}{s_{\beta}c_{\beta}} \right)$,  $m^2_{h^{\pm}_2}= (v^2 c_{\beta}^2+w^2)\left(\frac{\tilde{\lambda}_{23}}{2}  +ft_{\beta}\right)  $,  and $m^2_{G_{W}}=m^2_{G_{Y}}=0$. The original and mass eigenstates of the charged Higgs bosons are described as in Ref. \citep{Ninh:2005su} as
\begin{eqnarray}
	\left( \begin{array}{c}
		\eta^\pm\\
		\rho_1^\pm 
	\end{array} \right)=\begin{pmatrix}
		-s_{\beta}&  c_{\beta} \\
		c_{\beta}&  s_{\beta} \\
	\end{pmatrix} \left(\begin{array}{c}
		G_W^\pm
		\\ h_1^\pm
	\end{array} \right), \quad \left( \begin{array}{c}
		\rho_2^\pm \\
		\chi^\pm
	\end{array} \right) = \left( \begin{array}{cc}
		- s_{13} & c_{13} \\
		c_{13} & s_{13}
	\end{array} \right) \left(\begin{array}{c}
		G_Y^\pm
		\\ h_2^\pm
	\end{array} \right),
	\label{EchargedH}
\end{eqnarray}
where $t_{13}=v_1/w$, 
and 
\begin{align}\label{eq_fdepen}
	f&= \frac{c_{\beta } s_{\beta } \left(2 m_{h^\pm_1}^2 -\tilde{\lambda}_{12} v^2\right)}{2 \omega ^2}.
\end{align}
Three neutral gauge bosons are predicted by this model, where there is one zero eigenvalue corresponding to the massless photon.    Defining \cite{Buras:2012dp} 
\begin{align} \label{sc331}
s_{331}\equiv \sin\theta_{331}&=\sqrt{1- \frac{t_W^2}{3}}, \; 
c_{331}\equiv \cos\theta_{331} =-\frac{t_W }{\sqrt{3}},
\end{align}
The original and physical basis of the  neutral gauge bosons are presented below
\begin{align}
\begin{pmatrix}
X_\mu \\
W^3_\mu\\
W^8_\mu
\end{pmatrix} &= \begin{pmatrix}
s_{331} &0 & c_{331}  \\
0 & 1 & 0 \\
c_{331} & 0 & -s_{331}
\end{pmatrix}
\begin{pmatrix}
c_W & -s_W & 0 \\
s_W & c_W& 0 \\
0 & 0 & 1
\end{pmatrix}
\begin{pmatrix}
1 &   0 & 0\\
0 & c_\theta & -s_\theta \\
0 & s_\theta & c_\theta
\end{pmatrix}\begin{pmatrix}
A_\mu \\
Z_{1\mu} \\
Z_{2\mu}
\end{pmatrix}=C\begin{pmatrix}
A_\mu \\
Z_{1\mu} \\
Z_{2\mu}
\end{pmatrix},\crn 
C&=\begin{pmatrix}
s_{331}c_W &   \left(- s_{331}s_Wc_\theta+c_{331}s_\theta\right)  & \left( s_{331}s_Ws_\theta +c_{331}c_\theta\right) \\
s_W & c_Wc_\theta & -s_\theta c_w \\
c_{331}c_W &   -\left( c_{331}s_Wc_\theta+s_{331}s_\theta\right)  &\left(  c_{331}s_Ws_\theta -s_{331}c_\theta\right) 
\end{pmatrix}.
\label{neutralgaugebosonmix}
\end{align}
%
Next, the  gauge bosons are defined such as $Z_1\equiv Z$ and $Z_2\equiv Z'$, where $Z$ is the boson found experimentally.

The model includes five CP - odd neutral scalar components contained in the five neutral Higgs boson  $\eta^0_1=(v_2 +R_1 + i  I_1)/\sqrt{2}$, $ \rho^0=(v_1 +R_2 + i  I_2)/\sqrt{2}$,  $ \chi^0_2=(\omega +R_3 + i  I_3)/\sqrt{2}$, $ \eta^0_2 =(R_4 + i  I_4)/\sqrt{2}$,  and $ \chi^0_1=(R_5 + i  I_5)/\sqrt{2}$.  Which have three Goldstone bosons of  the neutral gauge bosons as $Z,Z'$, and $X^0$. Others are physical states with masses
\begin{align}
	m^2_{a_1}=\left(s_{\beta }^2 v^2+\omega ^2\right)  \left( f t^{-1}_{\beta} + \frac{1}{2}\tilde{\lambda}_{13} \right), \quad  m^2_{a_2}=f \left(\frac{\omega ^2}{c_{\beta }s_{\beta
	}}+c_{\beta } s_{\beta } v^2\right). 
\end{align} 
Consequently, {the condition $f > 0$ is} required for the term $f$. 

Examining the CP-even scalars, in two bases $(\eta^0_2,\; \chi^0_1)$ and $(\eta^0_1,\; \rho^0_1, \chi^0_1)$, {there} are corresponding $2\times2$ and $3\times 3$ sub-matrices for {the} masses of these Higgs boson, detailed
\begin{align}
	M^2_{0,3}&= \left(
	\begin{array}{ccc}
		\frac{c_{\beta } f \omega ^2}{s_{\beta }}+2 s_{\beta }^2 \lambda _1 v^2 & c_{\beta } s_{\beta } \lambda _{12} v^2-\omega ^2 f & \omega  (s_{\beta } \lambda _{13}-c_{\beta } f) v \\
		c_{\beta } s_{\beta } \lambda _{12} v^2-\omega ^2 f & \frac{s_{\beta } f \omega ^2}{c_{\beta }}+2v^2 c_{\beta }^2 \lambda _2  & v\omega  (c_{\beta } \lambda _{23}-s_{\beta } f)  \\
		 v \omega  (s_{\beta } \lambda _{13}-c_{\beta } f) & v\omega  (c_{\beta } \lambda _{23}-s_{\beta } f)  & 2 \lambda _3 \omega ^2+c_{\beta } s_{\beta } f v^2 \\
	\end{array}
	\right),
	\crn M^2_{0,2}&= \left(
	\begin{array}{cc}
		\frac{1}{2} \omega ^2 \left(\tilde{\lambda }_{13}+\frac{2 c_{\beta } f}{s_{\beta }}\right) & \frac{1}{2} v\omega  (\tilde{\lambda }_{13} s_{\beta }+2 c_{\beta } f)  \\
		\frac{1}{2} v \omega  (\tilde{\lambda }_{13} s_{\beta }+2 c_{\beta } f)  & \frac{1}{2} v^2s_{\beta } (\tilde{\lambda }_{13} s_{\beta }+2 c_{\beta } f)  \\
	\end{array}
	\right). 
\end{align}

The matrix $M^2_{0,2}$ has one zero value, and $m^2_{h_4}=\left(\frac{f}{t_{\beta }}+\frac{\tilde{\lambda }_{13}}{2}\right) \left(s_{\beta }^2 v^2+\omega ^2\right)$, consistent with the Goldstone boson of $X^0$ and the Higgs boson $h^0_4$ (heavy neutral boson) with mass at the $SU(3)_L$ breaking scale.  Furthermore, research also shows that Det$[\left.M^2_{0,3}]  \right|_{v=0}=0$, but Det$[ M^2_{0,3}] \neq0$. This implies that the electroweak scale needs to {have} at least one Higgs boson mass, which can be associated with the SM-like Higgs boson. Namely, it can be demonstrated that
\begin{align}\label{eq_Ch1}
	C^{h}_1	\left.M^2_{0,3}C^{hT}_1 \right|_{v=0}=\mathrm{diag}\left( 0,\; 2\lambda_3 w^2, fw^2/(s_\beta c_\beta)\right),\; C^{h}_1=\left(
	\begin{array}{ccc}
		s_{\beta } & c_{\beta } & 0 \\
		-c_{\beta } & s_{\beta } & 0 \\
		0 & 0 & 1 \\
	\end{array}
	\right),
\end{align}
and $C^{h}_1M^2_{0,3}C^{hT}_1\equiv\;M'^2_{0,3} $ satisfying
\begin{align}
	\label{eq_Mp203}	
	\left( M'^2_{0,3}\right)_{11}&= 2 v^2 \left(c_{\beta }^4 \lambda _2+c_{\beta }^2 \lambda _{12} s_{\beta }^2+\lambda _1 s_{\beta }^4\right),
	\crn 
	\left( M'^2_{0,3}\right)_{22}&=2v^2  s_{\beta }^2 c_{\beta }^2 (\lambda _1-\lambda _{12}+\lambda _2) +\frac{f \omega ^2}{c_{\beta } s_{\beta }},
	\crn 
	\left( M'^2_{0,3}\right)_{33}&= fv^2  s_{\beta } c_{\beta }+2 \lambda _3 \omega ^2,
	\crn 
	\left( M'^2_{0,3}\right)_{12}&=\left( M'^2_{0,3}\right)_{21}=v^2 s_{\beta }c_{\beta }  \left(s_{\beta }^2 (\lambda _{12}-2 \lambda _1)-c_{\beta }^2 (\lambda _{12}-2 \lambda _2)\right),
	\crn 
	\left( M'^2_{0,3}\right)_{13}&=\left( M'^2_{0,3}\right)_{31}= v \omega  \left(-2 f  s_{\beta }c_{\beta }+c_{\beta }^2 \lambda _{23}+\lambda _{13} s_{\beta }^2\right),
	\crn 
	\left( M'^2_{0,3}\right)_{32}&=\left( M'^2_{0,3}\right)_{23} =v \omega  \left(f c_{\beta }^2-f s_{\beta }^2+s_{\beta }c_{\beta }  (\lambda _{23}-\lambda _{13})\right).
\end{align}
As a result, by using unitary transformation $C^{h}_2$ with $\left( C^{h}_2\right)_{ij}\sim \mathcal{O}(v/w)$  ($i\neq j$) so that $C^{h}_2M'^2_{0,3}C^{hT}_2= \mathrm{diag}\left(m^2_{h^0_1},\; m^2_{h^0_2},\;m^2_{h^0_3} \right) $ and $m^2_{h^0_1}\sim \mathcal{O}(v^2)$  \cite{Nguyen:2018rlb}. Hence  $h^0_1\equiv h$ is determined with the SM-like Higgs boson found by experiment at the LHC. {Contributions to $h^0_1$ come only from Re$[\eta^0_1]=R_1/\sqrt{2}$ and Re$[\rho^0_1]\sim/\sqrt{2}$, in particular}
	\begin{equation}\label{eq_h01}
	R_1=s_{\beta} \times  h^0_1 -c_{\beta} \times h^0_2,\; R_2= c_{\beta} \times h^0_1 +s_{\beta} \times h^0_2. 
	\end{equation}
	Based on the presented result, the couplings and their one-loop contributions {to the aforementioned decays} will be derived and are shown in detail in the next section.

\section{\label{sec_coupling} Couplings and analytic formulas}
\begin{figure}[ht]
	\centering 
	\includegraphics[width=14cm]{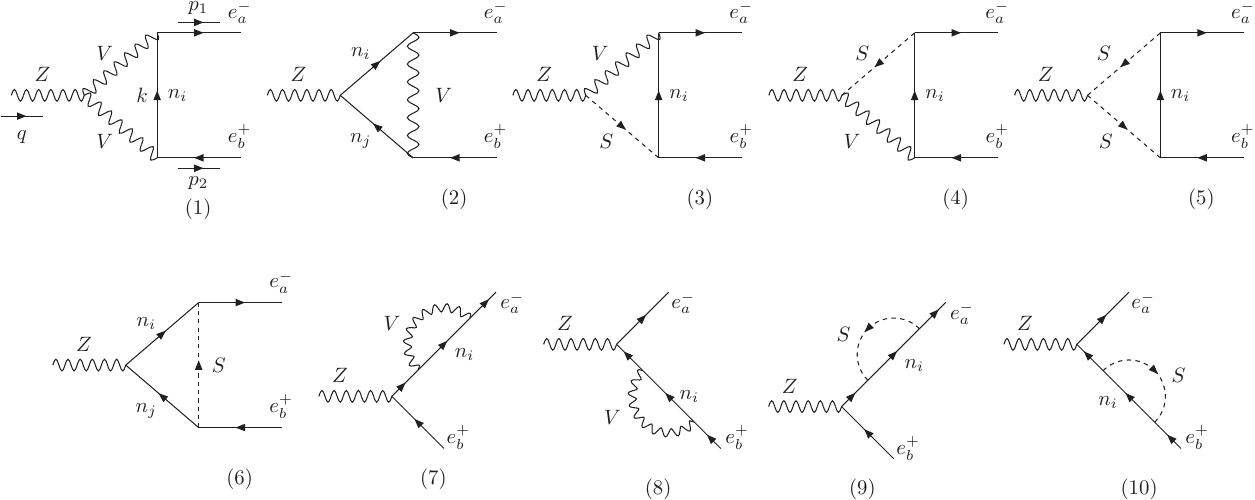}
	\caption{ One-loop Feynman diagrams contributing to $Z \to e_a e_b$ in the unitary gauge \cite{Jurciukonis:2021izn}}
	\label{Zeaeb}
\end{figure}

The effective amplitude for the decays  $Z(q)\to e^\pm_a (p_1) e^{\mp}_b (p_2)$ is {given by} \cite{Jurciukonis:2021izn, DeRomeri:2016gum}
\begin{align} \label{eq_Mzeab}
i\mathcal{M}(Z\to e^+_be^-_a)= & \frac{ie}{16\pi^2} \overline{u}_{a}\left[ \slashed{\varepsilon} \left( \bar{a}_l P_L + \bar{a}_r P_R\right) +  (p_1.\varepsilon) \left( \bar{b}_l P_L + \bar{b}_r P_R\right)  \right]v_{b},
\end{align}
where $\varepsilon_{\alpha}(q)$, $u_a(p_1)$, and $v_b(p_2)$ are the polarization of $Z$ and two Dirac spinors, respectively.  The massive gauge boson $Z$ has three physical polarization states satisfying  $q.\varepsilon=0$, which implies that  $p_2.\varepsilon=-p_1.\varepsilon$ with $q=p_1+p_2$.   Therefore,  Eq. \eqref{eq_Mzeab}  reduces {to} two dependent one-loop form factors proportional to $p_2.\varepsilon$ introduced in Ref. \cite{Jurciukonis:2021izn}. The correlating partial decay width is 
\begin{align} \label{}
\Gamma (Z\to e^+_b e^-_a)= 	\frac{\sqrt{\lambda}}{16\pi m_Z^3}\times \left(\frac{e}{16\pi^2}\right)^2 \left( \frac{\lambda M_0}{12 m^2_Z} +M_1 +\frac{\lambda M_2}{3 m^2_Z}\right),
\end{align}
where $\lambda= m^4_Z +m^4_b +m^4_a -2(m^2_Zm^2_a +m^2_Zm^2_b +m^2_am^2_b)\simeq m_Z^4$ because of $m^2_a,m^2_b\ll m^2_Z$, 
and 
\begin{align}
\label{eq_Mi}
M_0= & \left(m^2_Z -m^2_a -m^2_b\right) \left(|\bar{b}_l|^2 +|\bar{b}_r|^2\right)  -4 m_a m_b \mathrm{Re}\left[ \bar{b}_l  \bar{b}^*_r\right]\crn
-& 4m_b \mathrm{Re}\left[ \bar{a}^*_r \bar{b}^*_l   + \bar{a}^*_l \bar{b}^*_r  \right] -  4m_a\mathrm{Re}\left[ \bar{a}^*_l \bar{b}^*_l   + \bar{a}^*_r  \bar{b}^*_r  \right] , 
\crn M_1 = & 4 m_am_b \mathrm{Re}\left[\bar{a}_l\bar{a}_r^* \right],
\crn  M_2 = &  \left[ 2 m^4_Z - m_Z^2\left( m_a^2 + m_b^2\right) - \left( m_a^2 - m_b^2\right)^2  \right] \left( |\bar{a}_l|^2 +|\bar{a}_r|^2\right).
\end{align}
The relevant one-loop diagrams contributing to the decay amplitude $Z \to e_a e_b$ in the unitary gauge are illustrated in Fig. \ref{Zeaeb}.
The couplings of charged gauge bosons giving one-loop contributions to LFV$Z$ amplitudes are
\begin{align}
	\label{eq_g2Lagrangian}
	\mathcal{L}^Y_{\mathrm{V ff}} =& \frac{g}{\sqrt{2} }\sum_{a=1}^3 \sum_{i=1}^{9} \overline{n_i}\gamma^\mu P_L e_a \left[ 	U^{\nu*}_{ai}W_\mu^+ + U^{\nu*}_{(a+3)i}Y^+_\mu  \right] + h.c,
\end{align}

The couplings between gauge boson and two different leptons from Lagrangian in Eq.~(\ref{eq_g2Lagrangian}) are presented in Table~\ref{numbers}.
\begin{table}[h]
	\begin{center}
				\begin{tabular}{|c|c|}
			\hline
			Vertex & Coupling \\
			\hline
				$W^+_\mu \bar{n}_i e_b,  W^-_\mu \bar{e}_a n_i, $ & $\frac{ig}{\sqrt{2}}U^{\nu*}_{ai}\gamma^\mu P_L, \frac{ig}{\sqrt{2}}U^{\nu}_{ai} \gamma^\mu P_L $ \\
						\hline
			$ Y^+_\mu \bar{n}_i e_b, Y^-_\mu \bar{e}_a n_i $ & $\frac{ig}{\sqrt{2}}U^{\nu*}_{(a+3)i} \gamma^\mu P_L, \frac{ig}{\sqrt{2}}U^{\nu}_{(a+3)i} \gamma^\mu P_L$ \\
			\hline
							\end{tabular}
	\end{center}
\caption{\fontsize{13}{14}\selectfont~Feynman rules relevant one-loop contributions to $Z \to e_a e_b$ in the unitary gauge relate between gauge boson and two different leptons.  \label{numbers}}
\end{table}

The couplings between three gauge bosons originate from the covariant kinetic Lagrangian of the non-Abelian gauge bosons:
\begin{equation}\label{eq_LDg}
\mathcal{L}^g_D= -\frac{1}{4}\sum_{a=1}^8F^a_{\mu\nu}F^{a\mu\nu},
\end{equation} 
where 
\begin{equation}\label{eq_Famunu}
F^a_{\mu\nu}=\partial_{\mu}W^a_{\nu} -\partial_{\nu}W^a_{\mu} + g \sum_{b,c=1}^8f^{abc}W^b_{\mu} W^c_{\nu},
\end{equation}
where $f^{abc}$ $(a,b,c=1,2,...,8)$ are structure constants of the $SU(3)$ group, which identified as 
\begin{align}\label{eq_ZAgg}
\mathcal{L}^g_D =& -\sum_{V=W,Y}eg_{ZVV}Z^{\mu}(p_0)V^{+\nu}(p_+)V^{-\lambda}(p_-)\times \Gamma_{\mu\nu\lambda}(p_0,p_+,p_-), \crn 
&-e \sum_{V=W,Y} A^{\mu}(p_0)V^{+\nu}(p_+)V^{-\lambda}(p_-)\times \Gamma_{\mu\nu\lambda}(p_0,p_+,p_-) +\dots ,
\end{align}
where $\Gamma_{\alpha\beta\sigma}(p_0,p_+,p_-) \equiv g_{\alpha\beta}(p_0-p_+)_{\sigma} +g_{\beta\sigma}(p_+ -p_-)_{\alpha} +g_{\sigma\alpha}(p_--p_0)_{\beta}$, and $V=W^\pm,Y^\pm$.

In {T}able~\ref{table_3gaugcoupling} show the involving couplings of $Z$.
\begin{table}[ht]
	\centering \begin{tabular}{|c|c|}
		\hline
		Vertex & Coupling\\
		\hline 
		$g_{ZW^{+}W^{-}}$& $t_W^{-1}c_{\theta}
		$\\
		\hline 
		$g_{ZY^{+}Y^{-}}$& $ \frac{  1}{2s_W} \left[ c_{\theta} c_W \left(1	-  t^2_W\right) +s_{\theta} \sqrt{3 - t_W^2}\right] 
		$\\
		\hline 
	\end{tabular}
	\caption{Feynman rules for  triple gauge couplings corresponding with LFV$Z$ decays.}  \label{table_3gaugcoupling}
\end{table}

The covariant kinetic terms of the Higgs bosons, {which} include the couplings of Higgs and gauge bosons, are presented {as follows}
\begin{align}
	\label{eq_lkHiggs}
	\mathcal{L}^{\phi}_{\mathrm{kin}}=& \sum_{\phi=\chi, \rho, \eta}\left(D_{\mu}\phi \right)^{\dagger} \left(D^{\mu}\phi \right) \crn
	= &~ g_{hWW} g_{\mu\nu}hW^{-\mu} W^{+\nu} + g_{hYY} g_{\mu\nu}hY^{-\mu} Y^{+\nu}\crn 
	& + \left[ -ig^*_{hh^\pm_1W} \left( h^{+}_1\partial_{\mu}h -h\partial_{\mu}h^{+}_1 \right) W^{-\mu} + ig_{h^\pm_1W} \left( h^{-}_1\partial_{\mu}h -h\partial_{\mu}h_1^{-} \right)W^{+\mu} \right]\crn 
	&+\left[ -ig^*_{hh^\pm_2Y} \left( h^{+}_2 \partial_{\mu}h -h\partial_{\mu}h^{+}_2 \right) Y^{-\mu} + ig_{h^\pm_2Y} \left( h^{-}_2 \partial_{\mu}h -h\partial_{\mu}h_2^{-} \right)Y^{+\mu} \right]\crn 
	& +\sum_{k=1}^2 ie  A^{\mu}\left( h^{-}_k \partial_{\mu} h^{+}_k -h^{+}_k\partial_{\mu} h^{-}_k \right) + \sum_{k=1}^2 ieg_{Zh^+_kh^-_k}Z^{\mu} \left( h_k^{-}\partial_{\mu}h_k^{+} -h^{+ }_k\partial_{\mu}h_k^{-} \right)  
	\crn &
	+ Z^{\mu}e \left[ ig_{ZW^+h^-_1}  W^{+\nu}h_1^{-} g_{\mu\nu} + ig^*_{ZW^+h^-_1} W^{-\nu}h_1^{+} g_{\mu\nu}\right] 
	\crn &
	+ Z^{\mu}e \left[ ig_{ZY^+h^-_2}  Y^{+\nu}h_2^{-} g_{\mu\nu} + ig^*_{ZY^+h^-_2} Y^{-\nu}h_2^{+} g_{\mu\nu}\right] + \text{irrelevant terms}.  
\end{align}

{Next}, we present the corresponding terms contributing to the decays $Z\rightarrow e_ae_b$ from the second line in Eq.~(\ref{eq_lkHiggs}). Table~\ref{table_ZBB} show{s} the relevant Feynman rules, where $\partial_{\mu}h\rightarrow -ip_{0\mu}h$ and $\partial_{\mu}h_i^{\pm }\rightarrow -ip_{\pm\mu}h_i^{\pm }$, leading to $i\left( h^{-}_i\partial_{\mu} h^{+}_i -h^{+}_i\partial_{\mu} h^{-}_i \right)=h^-_ih^+_i(p_+-p_-)_{\mu}$. The incoming momenta are denoted $p_0$, $p_{\pm}$.
\begin{table}[ht]
	\centering 
	\begin{tabular}{|c|c|}
		\hline
		Vertex& Coupling\\
		\hline
		$g_{Zh^+_1h^-_1}$	& $ \frac{1}{2c_Ws_W}\left[c_{\theta } \left(1-2 s_W^2\right) +\frac{c_W s_{\theta } \left(2 \sqrt{3} c_{\beta }^2{-\frac{1}{\sqrt{3}}}   t_W^2-\sqrt{3}\right)}{3 \sqrt{1-{\frac{1}{{3}}} t_W^2}} \right] $\\
	\hline 	
		$g_{Zh^+_2h^{-}_2}$	& $\frac{1}{2s_Wc_W}\left[ c_{\theta } \left(s_{13}^2-2 s_W^2\right) -\frac{ s_{\theta } \left(c_{13}^2 -2\right)  \left(2 s_W^2-1\right)}{\sqrt{3-4 s_W^2}} \right]  $\\
		\hline 	
		$g_{ZW^{+}h^-_1}$& $-\frac{ s_{\theta } s_{2\beta } m_W  }{t_W \sqrt{3 -4 s_W^2}}$\\
			\hline 
		$g_{ZY^{+}h^{-}_2}$ & $-\frac{c_\beta c_{13} m_W  }{c_W s_W} \left[  c_{\theta } + \frac{ s_{\theta } \left(-1+2  s_W^2\right)}{ \sqrt{3-4 s_W^2}} \right] $\\
		\hline 
	\end{tabular}
	\caption{Feynman rules of couplings with $Z$ to charged Higgs and gauge bosons. }\label{table_ZBB}
\end{table}

The Lagrangian with $Z$ couplings to leptons {is given by:}
\begin{align}
	\label{eq:LZll}
	L_{Z\ell\ell}= & e \left( c_{\theta} + \frac{s_{\theta} }{\sqrt{3-4s_W^2}}\right) \overline{e_a}\gamma^{\mu} \left[\frac{1}{s_Wc_W}\left( -\frac{1}{2} +s_W^2 \right) P_L  +t_W P_R \right]e_a  Z_{1\mu}
	\crn &+  \frac{ec_{\theta }}{2 c_W s_W} \left[  \left(1 +\frac{t_{\theta } \left(2 s_W^2-1\right)}{ \sqrt{3-4 s_W^2}}\right)\overline{\nu_a}\gamma^{\mu}P_L  \nu_a + \frac{4t_{\theta } c_W^2}{ \sqrt{3-4 s_W^2}}\overline{N_a}\gamma^{\mu}P_L  N_a   \right]  Z_{1\mu}.
\end{align} 
{Returning} according to Feynman rules, {the} couplings {of the} $Z$ to two Majorana leptons $n_i$ and $n_j$ {can be expressed} using the form: 
\begin{align}
	\label{eq_LintM1}
	\mathcal{L}_{\mathrm{Z\ell \ell}}&= \sum_{i,j=1}^{9} \left[  \frac{e}{2}\overline{n_i}\gamma^{\mu}\left( G_{ij} P_L  -G_{ji} P_R\right) n_j Z_{\mu}\right],\crn
	G_{ij}&= {\frac{c_{\theta }}{2 c_W s_W}} \sum_{c=1}^3 \left[  \left(1 +\frac{t_{\theta } \left(2 s_W^2-1\right)}{ \sqrt{3-4 s_W^2}}\right)U^{\nu*}_{ci}U^{\nu}_{cj} + \frac{4 t_{\theta } c_W ^2}{ \sqrt{3-4 s_W^2}}U^{\nu*}_{(c+3)i}U^{\nu}_{(c+3)j} \right]. 
\end{align}
Therefore, we  consider the limit $\theta=0$, {which leads} to the same coupling of $Z$ {as} given in Ref. \cite{Hong:2024yhk}, consistent with 2HDM results. 
\begin{align}
	\label{eq_LintM}
	\mathcal{L}_{\mathrm{int}}= \sum_{i,j=1}^{9} \left[ -\frac{g}{4m_W} \overline{n_i} \left( \lambda^h_{ij} P_L  + \lambda^{h*}_{ij} P_R\right) n_j h  + \frac{e}{2}\overline{n_i}\gamma^{\mu}\left( G_{ij} P_L  -G_{ji} P_R\right) n_j Z_{\mu}\right],
\end{align}
where 
\begin{align}
	\label{eq:Gij}
	\lambda^h_{ij}=& \lambda^h_{ji} = \sum_{c=1}^3 \left( U^{\nu*}_{ci}m_{n_i}U^{\nu}_{cj} + U^{\nu*}_{cj}m_{n_j}U^{\nu}_{ci}\right). 
\end{align}

Following Ref. \cite{Hong:2022xjg}, the Yukawa couplings of leptons  with charged Higgs boson are identified as
\begin{align}
	\label{eq_LHee}
	\mathcal{L}^{\ell n h^\pm}=- \frac{g}{\sqrt{2}m_W} \sum_{k=1}^2 \sum_{a=1}^3\sum_{i=1}^9 h_k^+\overline{n_i} \left(\lambda^{L,k}_{ai}P_L+\lambda^{R,k}_{ai}P_R\right)e_a   +\mathrm{h.c.},
\end{align} 
we defined
\begin{align}
	\lambda^{R,1}_{ai}&=m_{a}  t_{\beta}U^{\nu*}_{ai},  	\quad  \lambda^{L,1}_{ai} = s_{\beta}z_0e^{i\alpha_{23}}\sum_{c=1}^3(\tilde{m}_D)_{ac}U^{\nu}_{(c+3)i},
	\crn 	\lambda^{R,2}_{ai}&=\frac{m_{a}c_{\theta}U^{\nu*}_{(a+3)i}}{c_{\beta}}, \quad \lambda^{L,2}_{ai} = c_{\theta}z_0 \sum_{c=1}^3 \left[ -e^{i\alpha_{23}} (\tilde{m}_D)_{ac}U^{\nu}_{ci} + t^2_{\theta}(\tilde{M}^T_R)_{ac}U^{\nu}_{(c+6)i}\right] .
	\label{eq_lambdaLR}
\end{align}

Additionally, one-loop contributions of the decays $h \to e_a e_b$ were introduced previously, see Ref. \cite{Hong:2022xjg}, which will be used to investigate simultaneously with LFV$Z$  and cLFV decays  in this work.  We will use the analytical formulas given in Ref. \cite{Hong:2022xjg}.

\section{\label{sec_numerical} Numerical discussion}

In this paper, we will use the neutrino oscillation data {provided} in Refs. \cite{ParticleDataGroup:2020ssz, T2K:2019bcf}. The lepton mixing matrix $U_{\mathrm{PMNS}}$ has the standard form {as} a function of parameters defined from the experimental data, namely the three mixing angles $\theta_{ij}$ \cite{ParticleDataGroup:2020ssz, T2K:2019bcf,ParticleDataGroup:2018ovx}, one Dirac phase $\delta$ and two Majorana phases $\alpha_{1}$ and $\alpha_2$~\cite{ParticleDataGroup:2018ovx}, in particular
\begin{align}
	U^{\mathrm{PDG}}_{\mathrm{PMNS}}&=f(\theta_{12},\theta_{13},\theta_{23},\delta)\times  \;\mathrm{diag}\left(1, e^{i\alpha_{1}},\,e^{i\alpha_{2}}\right),
\crn f(\theta_{12},\theta_{13},\theta_{23},\delta)&\equiv  \begin{pmatrix}
		1	& 0 &0  \\
		0	&c_{23}  &s_{23}  \\
		0&  	-s_{23}& c_{23}
	\end{pmatrix}\,\begin{pmatrix}
		c_{13}	& 0 &s_{13}e^{-i\delta}  \\
		0	&1  &0  \\
		-s_{13}e^{i\delta}&  0& c_{13}
	\end{pmatrix}\,\begin{pmatrix}
		c_{12}	& s_{12} &0  \\
		-s_{12}	&c_{12}  &0  \\
		0& 0 	&1
	\end{pmatrix},  \label{eq_UnuPDG}
\end{align}
where $c_{ij}\equiv\cos\theta_{ij}$, $s_{ij}\equiv\sin\theta_{ij}$, $i,j=1,2,3$ ($i<j$), $0\le \theta_{ij}<90\; [\mathrm{Deg.}]$ and $0<\delta\le 720\;[\mathrm{Deg.}]$. We fix the range $-180\le\alpha_i\le 180$ [Deg.] for the Majorana phases.  For numerical research, benchmark {related} to the NO (normal order) of the neutrino oscillation data \cite{ParticleDataGroup:2022pth} will be selected as the input to fix $\tilde{m}_D$. {Specifically, these parameters are:} $s^2_{12}=0.32$, $s^2_{13}= 0.0216$, $s^2_{23}= 0.547$,  $ \Delta m^2_{32}=2.424\times 10^{-3} [\mathrm{eV}^2]$, $\Delta m^2_{21}=7.55\times 10^{-5} [\mathrm{eV}^2]$, $\delta= 180 \;[\mathrm{Deg}] $, and $\alpha_1=\alpha_2=0$. As {a} consequence, the matrix $\tilde{m}_D$, which the reduced Dirac mass matrix, is chosen below
\begin{equation}\label{eq_tmD}
\tilde{m}_D=\begin{pmatrix}
0	&0.613  & 0.357 \\
-0.613	& 0 & 1 \\
-0.357	&-1  &0 
\end{pmatrix}.
\end{equation}
The mixing matrix $V_R$ is parameterized {similarly} as  $V_{R}= f(\theta^r_{12},\theta^r_{13},\theta^r_{23},0)$, where the scanning ranges for the angles $\theta^r_{ij} \in \left[ 0,2\pi\right]$. The remaining free terms are explored within the following ranges:
\begin{align} \label{eq_scanRange}
\hat{k}_{1,2,3}&\geq 5,\; 1 \;\; [\mathrm{TeV}]\leq m_{h^\pm_{1}}, m_{h^\pm_{2} }\leq 5\; \; [\mathrm{TeV}], 
\crn t_{\beta}&\in [0.5,40],\; 100\; [\mathrm{GeV}]\leq z\leq 600 \; [\mathrm{GeV}].
\end{align}
$G_F=1.663787 \times 10^{-5}\;[\mathrm{GeV}^{-2}]$,  $\alpha_{em}=e^2/(4\pi)=1/137$, $g=0.652$, $s^2_W=0.231$, $m_e=5.0 \times 10^{-4}$ [GeV], $m_{\mu}=0.105$ [GeV], $m_{\tau}=1.776$ [GeV],  $m_Z=91.1876$ [GeV], and  $m_W=80.385$ [GeV], and the total decay width of the $Z$ boson is $\Gamma_Z= 2.4955$ [GeV]. {These parameters are found experimentally} \cite{ParticleDataGroup:2022pth}. The well-known decays {branching ratios} used in this research are Br$(\mu\to e\overline{\nu_e}\nu_{\mu})\simeq 1$,  Br$(\tau\to \mu\overline{\nu_{\mu}}\nu_{\tau})\simeq 0.1739$, and   Br$(\tau\to e\overline{\nu_e}\nu_{\tau})\simeq 0.1782$ \cite{ParticleDataGroup:2022pth}. The scanning ranges of $\hat{k}_{1,2,3}$  are suggested {based on} the numerical checks in Refs. \cite{Hue:2021xap, Hue:2020wnn}, {where} the values of $k_{ij}$ given in Eq. \eqref{eq:kij} must satisfy the ISS relation, i.e., $|k_{ij}| \simeq   |M_R|/|m_D|\gg 1 \forall i,j=1,2,3$. Note that {the} values of $k_{ij}$ {relate} to $\hat{k}_{1,2,3}$ through Eq. \eqref{eq_MRd}.

\begin{figure}[ht]
	\centering
	\begin{tabular}{ccc}
		&\includegraphics[width=5.6cm]{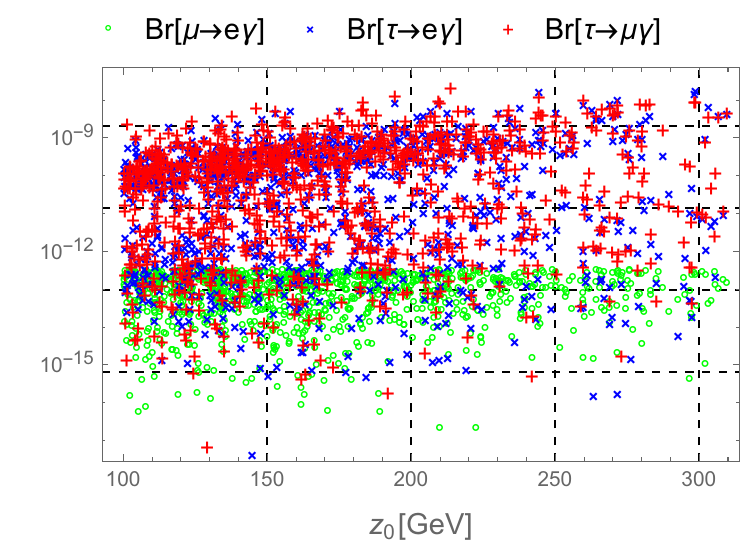}
	\includegraphics[width=5.6cm]{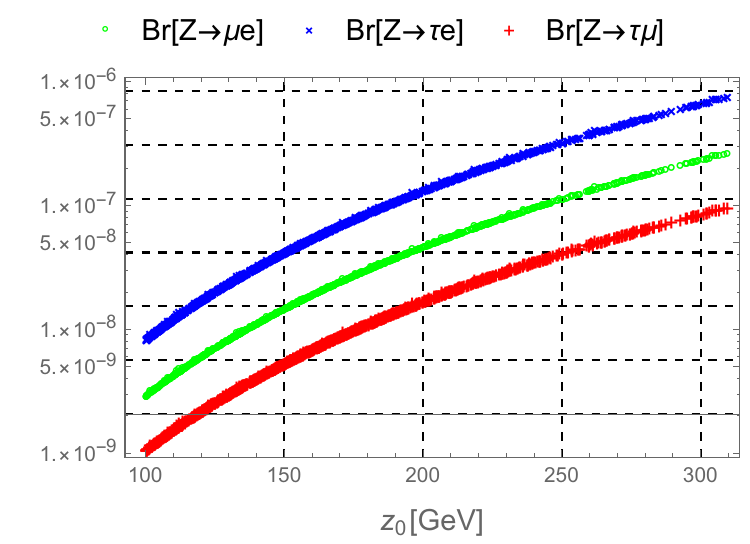} 
		&
	\includegraphics[width=5.6cm]{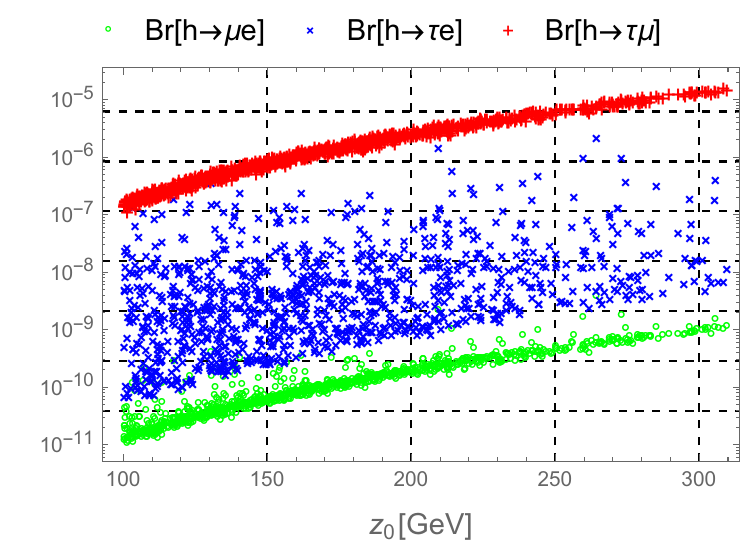} 
	\end{tabular}
	\caption{The dependence of  LFV decay rates on $z_0$.}\label{fig_z0LFV}
\end{figure}

In the numerical investigation, all collected points presented in the following figures satisfy current experimental constraints on the decay rates of cLFV, LFV$h$, and LFV$Z$ decays. The dependence of the LFV decay rates on $z_0$  is presented in Fig. \ref{fig_z0LFV}. The upper bounds of cLFV decays are consistent with {current experimental constraints, as shown in} Refs. \cite{BaBar:2009hkt, MEG:2016leq, Belle:2021ysv, MEGII:2023ltw} {and detailed in} Table \ref{BRdecay}. The future sensitivity  of Br($\mu \to e \gamma$), Br($\tau \to \mu \gamma$), and Br($\tau \to e \gamma$) may {reach} orders of $\mathcal{O}(10^{-14})$, $\mathcal{O}(10^{-9})$, and $\mathcal{O}(10^{-9})$, respectively, as {shown} in Table I \cite{MEGII:2018kmf, Belle-II:2018jsg}. Therefore, the upper bounds on {the} branching ratios of these decays will be larger than the future sensitivity.
 
According to illustrated in Fig. \ref{fig_z0LFV}, we can see that LFV$Z$ decay rates {strongly depend} on $z_0$, {shown} in three narrow curves in the middle panel of Fig. \ref{fig_z0LFV}. Additionally, an interesting consequence in the 331$NL$ model {is} that the upper constraint of Br$(Z\to \mu^+e^-)\leq 2.6\times 10^{-7}$ {is consistent} with {the} experimental {data} in Table~\ref{BRdecay}. The two LFV$h$ decays, $h\to \mu e$ {and} $h\to \tau \mu$, have upper bounds of Br$(h \to \mu e) < 5 \times 10^{-9}$ and  Br$(h \to \tau \mu) < 2 \times 10^{-5}$, {providing} the {strictest} upper bound  on $z_0$, namely $z_0\leq 310$ GeV.
Therefore, if one of these decay rates is detected, the remaining values can be predicted precisely. 

\begin{figure}[ht]
	\centering
	\begin{tabular}{ccc}
		\includegraphics[width=5.6cm]{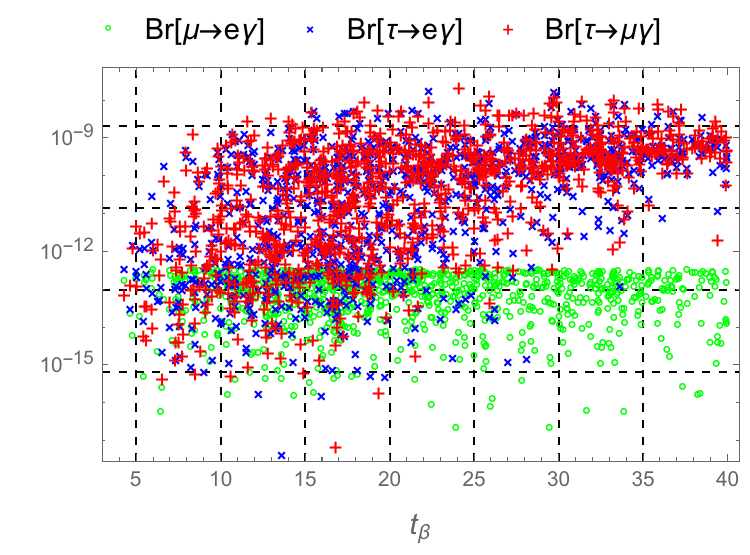}
		& \; 
		\includegraphics[width=5.6cm]{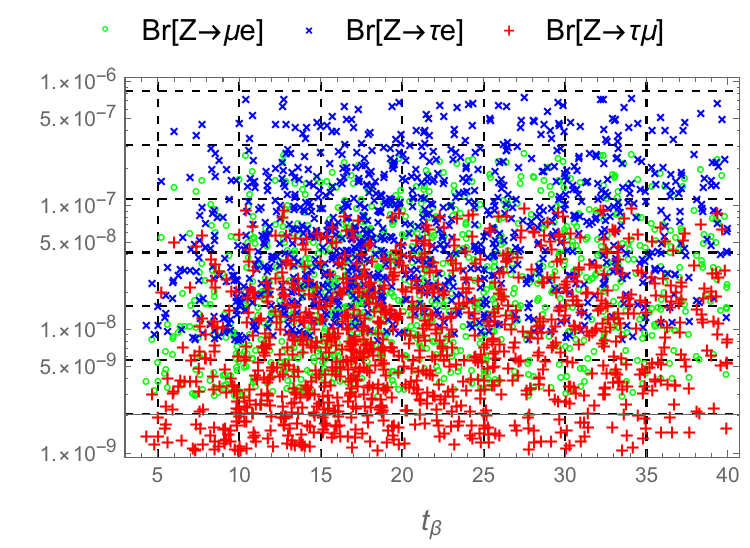} 
		&
		\includegraphics[width=5.6cm]{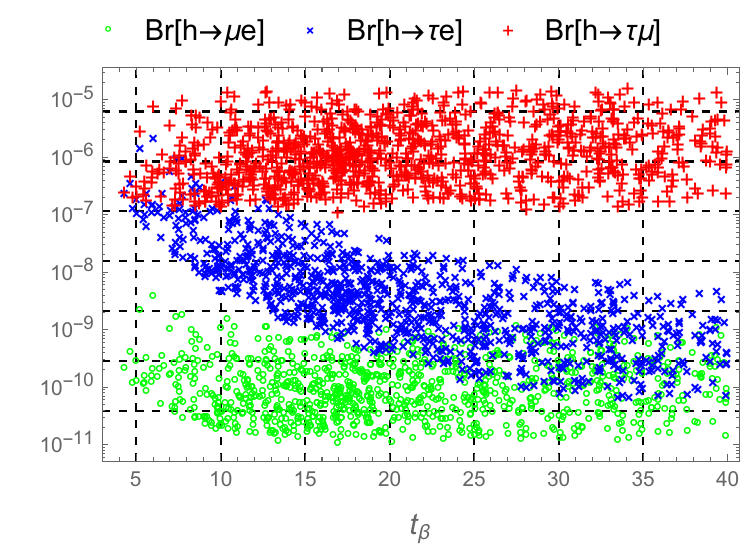}
	\end{tabular}
	\caption{ The dependence of  LFV decay rates on {$t_\beta$}.}\label{fig_tbLFV}

\end{figure}

Similarly, the dependence of LFV decays on $t_\beta$ is shown in  Fig. \ref{fig_tbLFV}. 
 Corresponding to the chosen scanning range of the parameter $t_\beta$ given in Eq. \eqref{eq_scanRange},  we see that all Br(LFV) depend on $t_\beta$, and the upper bounds are consistent with the constraints from current experiments as {shown} in Refs. \cite{BaBar:2009hkt, MEG:2016leq, Belle:2021ysv, MEGII:2023ltw}. However, the upper bounds on branching ratios in the left panel of Fig. \ref{fig_tbLFV} are larger than the future sensitivity, {as demonstrated in the} results in Fig. \ref{fig_z0LFV} \cite{MEGII:2018kmf, Belle-II:2018jsg}, see Table \ref{BRdecay}. Next, the upper bound of LFV$Z$ decays are max[Br$(Z \to \mu e)] \simeq 2.4 \times 10^{-7}$, max[Br$(Z \to \tau \mu)] \simeq 1. \times 10^{-7}$, and max[Br$(Z \to \tau e)] \simeq 8.8 \times 10^{-7}$. We can see that only the decay rate $(Z \to \mu e)$  is close to the latest experimental
		bound \cite{ATLAS:2022uhq}, {with} values of $\mathcal{O}(10^{-7}).$ Besides that, the two other decays are still much
		smaller than the recent experimental sensitivities as in Ref. \cite{ATLAS:2021bdj}, {with} values of $\mathcal{O}(10^{-6}).$ 
		However, the future sensitivities for these decays will be as at HL-LHC \cite{Dam:2018rfz} and at FCC-ee \cite{Dam:2018rfz, FCC:2018byv}, see Table \ref{BRdecay}. From these results, we can infer that although Br($Z \to \mu e$) satisfied the recent experimental sensitivities, two other channels satisfied the future experimental data at HL-LHC \cite{Dam:2018rfz}. However, all of them are still very difficult to detect at FCC-ee \cite{Dam:2018rfz, FCC:2018byv}. In addition, the upper bounds of Br$(h \to \mu e),$ Br$(h \to \tau \mu)$ decays have the same result as in Fig. \ref{fig_z0LFV}, and Br$(h \to \tau e) < 5 \times 10^{-6}$. The future experimental sensitivities at the HL-LHC and $e^+e^-$ colliders may be {of the order} $\mathcal{O}(10^{-5}), \mathcal{O}(10^{-4}),$ and $\mathcal{O}(10^{-4})$ \cite{Qin:2017aju, Barman:2022iwj, Aoki:2023wfb} for the three {mentioned previously} LFV$h$ decays, respectively. It is {easily seen} that only Br$(h \to \tau \mu)$ {is likely to} reach a significant value {within} the near future experimental sensitivities at the HL-LHC and $e^+e^-$ colliders.

\begin{figure}[ht]
	\centering
	\begin{tabular}{ccc}
		\includegraphics[width=5.6cm]{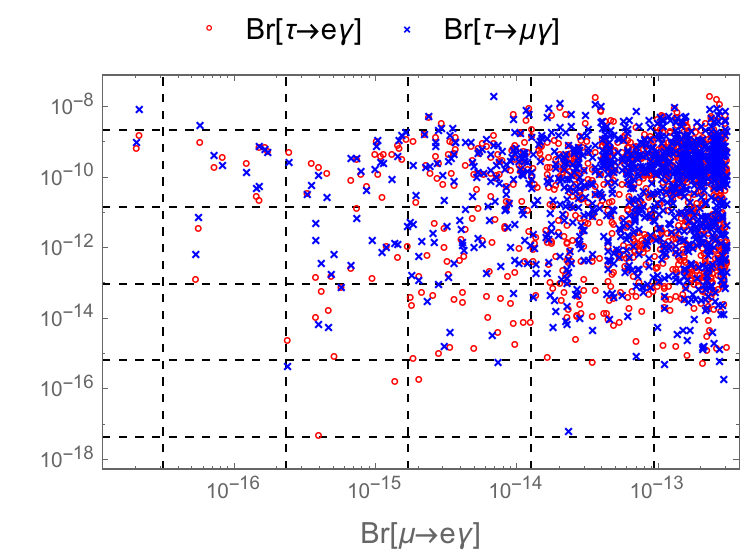}
		&
		\includegraphics[width=5.6cm]{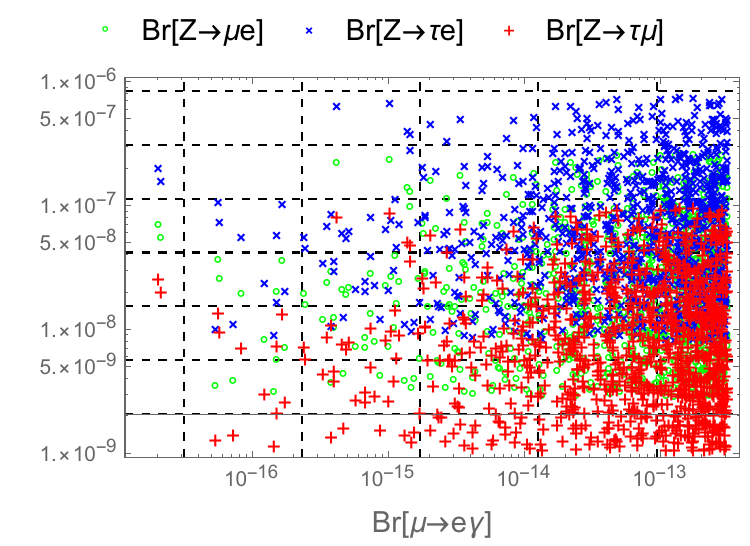} 
				\includegraphics[width=5.6cm]{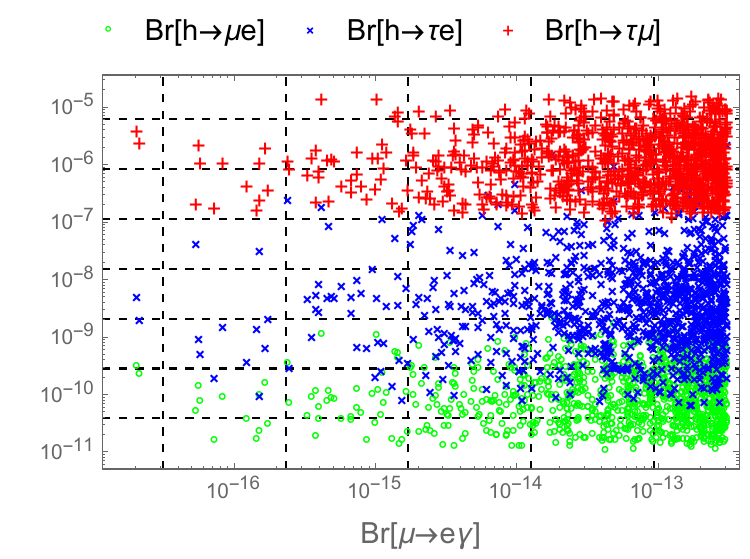}
	\end{tabular}
	\caption{ The dependence of  LFV decay rates on Br$(\mu \to e \gamma)$.}\label{fig_megaLFV}
\end{figure}

The correlations between LFV decays {and} Br$(\mu \to e \gamma)$ {are illustrated} in Fig. \ref{fig_megaLFV}. We choose a survey value of Br$(\mu \to e \gamma)$ that satisfies the experimental limits as in Ref. \cite{MEG:2016leq}. Satisfying the requirement of Br$(e_b \to e_a\gamma)$ is the most stringent requirement in experiments. Therefore, all investigations of decay processes should satisfy this condition. The decay rate Br$(\mu \to e \gamma)$ is the most stringently constrained by experiments. Future sensitivities of $\mathcal{O}(10^{-9})$ continue to support all other decay rates reaching their respective expected sensitivities \cite{Hong:2023rhg}, especially {with} the upper bound of these decays {being} Br$(\tau \to e(\mu) \gamma) \sim 10^{-8}$. 
{Our results show that} all decay rates can reach the corresponding expected future sensitivities, except {for} Br$(h \to \mu e)$, {which remains} smaller than future experimental sensitivities.

\begin{figure}[ht]
	\centering
	\begin{tabular}{ccc}
		\includegraphics[width=5.6cm]{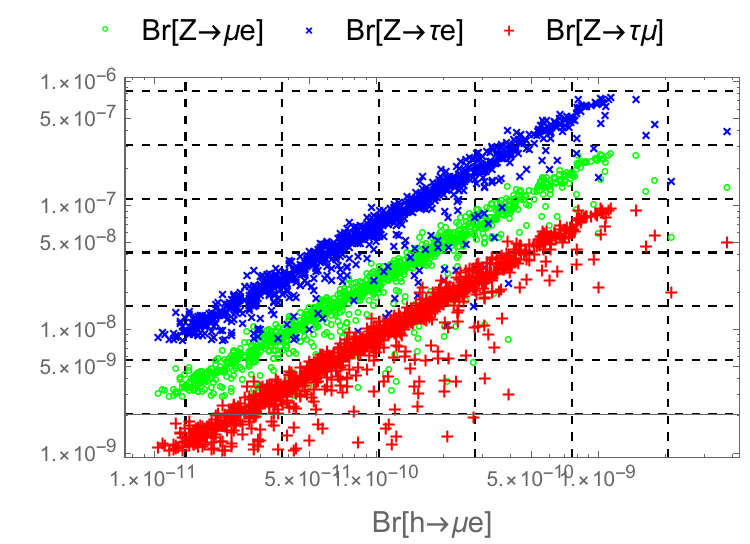}
		&
		\includegraphics[width=5.6cm]{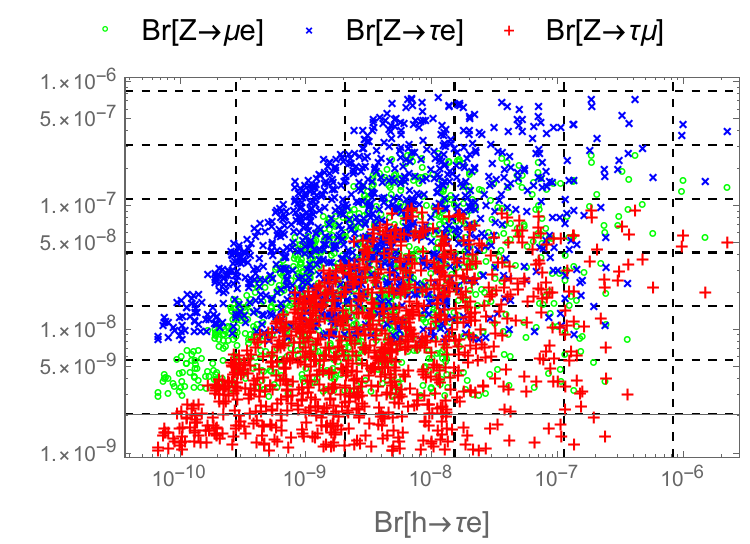} &
		\includegraphics[width=5.6cm]{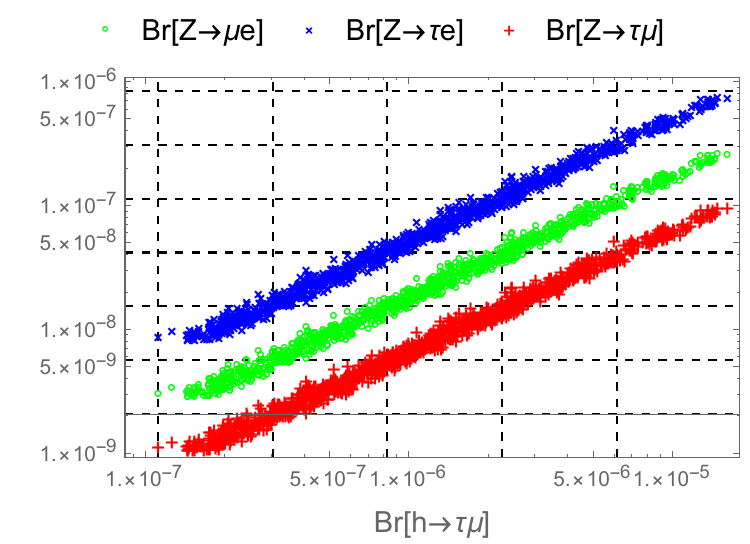}
	\end{tabular}
	\caption{The correlations between LFV$h$ and LFV$Z$ decay rates }\label{fig_LFVhz}
\end{figure}

The correlations between LFV$h$ and LFV$Z$ decay rates are presented in Fig. \ref{fig_LFVhz}.  Over the value range of Br$(h \to e_ae_b)$ that satisfies experimental constraints, all {branching ratios} of LFV$Z$ decay channels {strongly depend} on these values, as {shown} in Fig. \ref{fig_LFVhz}. Particularly, {a} large value of Br$(h \to e_a e_b)$ predicts {a correspondingly} large value of Br$(Z \to e_ae_b)$. The upper bounds of the branching ratio{s} of these decay are {close} to the recent experimental data. 

\section{\label{conclusion} Conclusion}
In the {331$NL$} model  framework, we have studied the cLFV, LFV$h$, and LFV$Z$ decays in detail and obtained many interesting results.  Firstly, the numerical results showed {a} strong dependence of LFV$Z$ decay on $z_0$. {In contrast, the dependencies of LFV decays} on $t_\beta$ are weaker. Only the decay rate $(Z \to \mu e)$  is close to the latest experimental bound, and Br$(h \to \tau \mu)$ can reach the near future experimental sensitivities. Secondly,  Br$(\mu \to e \gamma)$ is significantly proportional to Br$(Z \to \tau e)$ and Br$(h \to\tau \mu)$. All decay rates  can reach the corresponding expected future sensitivities, except {for} Br$(h \to \mu e)$.  Lastly,  Br$(h \to \mu e, \tau \mu)$ {exhibit} nearly linear  correlation with  Br$(Z \to e_ae_b)$. These clear dependencies indicate that if one of the two LFV$Z$ or LFV$h$ {is} observed by future experiments, the 331$NL$ {model} will predict the remaining decay rates. This model will also be confirmed or excluded if both LFV decay rates are measured experimentally. The upper bounds of Br$(Z \to e_ae_b)$ also coincide with recent experiments and consistent with previous discussions \cite{Hong:2023rhg}. In conclusion, the simultaneous study of the three LFV decay {types} in our work shows many interesting correlations that can be tested {in} upcoming experiments. {Furthermore}, the survey results are more reliable and our work helps {narrow down} the allowed regions of the parameter space that can be effectively probed by forthcoming experiments for LFV decays.

\section*{Acknowledgments}
This research is funded by An Giang University under grant number 23.01.CN.

\appendix

\section{\label{app_lfvh}One-loop form factors  for  LFV$Z$ decays the unitary gauge}
The form factors are calculated in the unitary gauge as in Ref. \cite{Hong:2024yhk} as below. 
Diagram (1) in Fig \ref{Zeaeb} give one-loop contributions: 
\begin{align}
	\bar{a}^{nGG}_l	=&   \sum_{i=1}^9 g^{LL}_{ab}g_{ZGG}
	\left\{ \left[ -4+  \dfrac{m^2_{n_i}}{m^2_G} \left(\frac{m_Z^2}{m_G^2} -2\right)\right] C_{00}  +2\left(m^2_Z- m^2_a -m^2_b\right)X_3 
	\right.\crn &\left. \hspace{3.7cm}
	-\frac{1}{m_G^2}  \left[\frac{}{} m^2_Z  \left(2m^2_{n_i} C_0 +m^2_aC_1 +m^2_bC_2\right)
	\right.\right.\crn	& \left.\left. 
	\hspace{5cm} -m^2_{n_i} \left(B_0^{(1)} +B_0^{(2)}\right) -m^2_aB_1^{(1)} -m^2_bB_1^{(2)}   \right]  \right\} , \label{eq_al1u} 
	%
	%
	\\	\bar{a}^{nGG}_ r=&    \sum_{i=1}^9 g^{LL}_{ab} g_{ZGG}m_am_b \left[  \left(-4 + \frac{ m^2_Z}{m_G^2} \right) X_3 
	+ \dfrac{m^2_Z -2 m_G^2}{m^4_G}C_{00} \right], \label{eq_ar1u}
	%
	\\	\bar{b}^{nGG}_l= &  \sum_{i=1}^9 g^{LL}_{ab}  g_{ZGG}m_a \left[ 4 \left( X_3 -X_1\right) 	+\dfrac{m^2_Z -2m_G^2}{m^4_G}   \left( m^2_{n_i} X_{01} +m^2_bX_2 \right)   -\frac{2m_Z^2}{m_G^2} C_2  
	\right], \label{eq_bl1u} 
	%
	\\	\bar{b}^{nGG}_r =&  \sum_{i=1}^9 g^{LL}_{ab}  g_{ZGG}m_b \left[4\left(X_3   - X_2\right)  +\dfrac{m^2_Z -2 m_G^2}{m^4_G}\left( m^2_{n_i}X_{02} +m^2_aX_1\right) -\frac{2m^2_Z}{m_G^2} C_1 \right], \label{eq_br1u}
\end{align}
where  $g^{LL}_{ab}\equiv g^{L*}_{aiG} g^{L }_{biG}$, the PV-functions  $B^{(k)}_{0,1}=B^{(k)}_{0,1}(p_k^2; m_{n_i}^2,m_G^2)$, $C_{00,0,k,kl}=C_{00,0,k,kl}(m_a^2,m_Z^2,m_b^2; m_{n_i}^2, m_G^2,m_G^2)$, and  $X_{0,k,kl}$  are identified in terms of the PV-functions  for all $k,l=1,2$. 

One-loop form factors from diagram {(2)}    in Fig. 1 are: 
\begin{align}
	\bar{a}^{Gnn}_l =&    \sum_{i,j=1}^9 \frac{g^{LL}_{ab}}{m_G^2} \left\lbrace G_{ij} \left[  m^2_G \Big(4C_{00}  +2 m^2_aX_{01} +2m^2_bX_{02} -2m^2_Z \left( C_{12}+X_0\right)\Big) \right.\right.\crn
	&\left.\left. \hspace{3.4cm} -\left(m_{n_i}^2 -m^2_a\right)B_0^{(1)} -\left(m_{n_j}^2-m^2_b\right)B_0^{(2)}  +m^2_aB_1^{(1)} +m^2_bB_1^{(2)}
	\right.\right.\crn
	&\left.\left. \hspace{3.4cm}+ \left(m_{n_j}^2m_{a}^2 +m_{n_i}^2m_{b}^2-m_a^2m_b^2 \right) X_0  -m_{n_i}^2m_{n_j}^2C_0 
	\right.\right.\crn
	&\left.\left. \hspace{3.4cm} -m_{n_i}^2m_{b}^2 C_1 -m_{n_j}^2m_{a}^2 C_2  \frac{}{}\right] \right.\crn
	&\left. \hspace{2.6cm}+G_{ji} m_{n_i}m_{n_j} \bigg[2m^2_G C_0 -2C_{00} -m^2_aC_{11} -m^2_bC_{22} 
	\right.\crn &\left. \hspace{5cm}	
	+\left(m^2_Z -m^2_a -m^2_b\right)C_{12}\bigg]\right\rbrace,  \label{eq_al2u} 
	%
	\\ \bar{a}^{Gnn}_r  =&  
	\sum_{i,j=1}^9 {\frac{g^{LL}_{ab}}{m_G^2}m_am_b} G_{ij} \left[\frac{}{} 2C_{00}+ 2m_{G}^2X_0 +m^2_aX_1 +m^2_bX_2 
	%
	 -m_Z^2 C_{12}  -m_{n_i}^2C_1 -m_{n_j}^2C_2\right], \label{eq_ar2u}
	\\ \bar{b}^{Gnn}_l  =&\sum_{i,j=1}^9  \frac{g^{LL}_{ab}}{m_G^2}{m_a}  \left[ G_{ij} \left(-2m_{G}^2 X_{01}   -m^2_bX_2 +m_{n_j}^2C_2\right) +G_{ji} m_{n_i}m_{n_j}(X_1 -C_1) \right],   \label{eq_bl2u} 
	%
	\\ \bar{b}^{Gnn}_r =& \sum_{i,j=1}^9  \frac{g^{LL}_{ab}}{m_G^2}{m_b}   \Big[G_{ij} \left( -2m_{G}^2X_{02}  -m^2_aX_1  +m_{n_i}^2C_1\right) +G_{ji} m_{n_i}m_{n_j} (X_2 -C_2)
	\Big], \label{eq_br2u}
\end{align} 
where  $g^{LL}_{ab}\equiv g^{L*}_{aiG} g^{L}_{bjG}$,  
$$B^{(1)}_{0,1}=B^{(1)}_{0,1}(p_1^2; m_{G}^2, m_{n_i}^2),  B^{(2)}_{0,1}=B^{(2)}_{0,1}(p_2^2; m_{G}^2, m_{n_j}^2), $$
and $C_{00,0,k,kl}=C_{00,0,k,kl}(m_a^2,m_Z^2,m_b^2;  m_{G}^2,m_{n_i}^2,m_{n_j}^2)$   for $k,l=1,2$.

Two diagrams (7) and (8) in Fig. \ref{Zeaeb} give sum contributions that have form factors 
\begin{align}
	\bar{a}^{nG}_l =&  \dfrac{ g^L_{Ze}}{m^2_G (m^2_a -m^2_b)} \sum_{i=1}^9 g^{LL}_{ab} 
	\left\lbrace 2m^2_{n_i}(m^2_a B_0^{(1)}- m^2_b B_0^{(2)})  + m^4_a B_1^{(1)}  -m^4_b B_1^{(2)}
	\right.\crn&\left.\hspace{5.5cm}+  \left(2 {m^2_G} +m^2_{n_i} \right) \left(m^2_aB_1^{(1)} -m^2_bB_1^{(2)} \right)  \right\rbrace. \label{eq_al3u} 
	%
	\\ \bar{a}^{nG}_ r =& \dfrac{ m_am_b g^R_{Ze}}{m^2_G (m^2_a -m^2_b)} \sum_{i=1}^9   g^{LL}_{ab} \left\lbrace 2m^2_{n_i}(B_0^{(1)}- B_0^{(2)})  + m^2_aB_1^{(1)}  -m^2_b B_1^{(2)}
	\right.\crn&\left. \hspace{5.5cm} +  \left(2m^2_G +m^2_{n_i}\right) \left(B_1^{(1)} -B_1^{(2)} \right)  \right\rbrace , \label{eq_ar3u}
	\\ \bar{b}^{nG}_{l} =&    \bar{b}^{n G}_{r} = 0, \label{eq_blr3u}
\end{align}
where  $g^{XY}_{ab}\equiv g^{X*}_{aiG} g^{Y }_{biG}$ with $X,Y=L,R$, and  $B^{(k)}_{0,1}=B^{(k)}_{0,1}(p_k^2; m_{n_i}^2,m_G^2)$ with $k=1,2$. 

In the limit $\theta=0$, the contributions from $W^\pm$ corresponding to diagrams (1), {(2)}, (7), and (8)  of Fig. \ref{Zeaeb} are:
\begin{align}
\label{eq:Wnx}
m_G&=m_W,\; g_{ZGG}=g_{ZWW}=t^{-1}_W,\; g^L_{aiG}=g^L_{aiW}= \frac{g}{\sqrt{2}} U^{\nu *}_{ai},\;
\end{align}
and $G_{ij}$ is given in Eq. \eqref{eq_LintM1}. 

The contributions from $Y^\pm$ corresponding to diagrams (1), {(2)}, (7), and (8)  of Fig. \ref{Zeaeb} are:
\begin{align}
\label{eq:Ynx}
 m_G&=m_Y,\; g_{ZGG}=g_{ZYY}= \frac{1-t_W^2}{t_W},\; g^L_{aiG}=g^L_{aiY}= \frac{g}{\sqrt{2}} U^{\nu *}_{(a+3)i},\;
\end{align}

The two diagrams {(3) and (4)} appearing in the model under consideration were not discussed previously, we list here for completeness the form factors of diagram {(3)} in Fig. 1 are:
\begin{align}
	\label{eq:abFVH}
	\overline{a}^{nGH}_l &= \sum_{i=1}^9 -\frac{g_{ZGH}  g^{L*}_{aiG}}{m_G^2} \left[ g^{L}_{biH}m_{n_i}\left(m_G^2C_0  -C_{00}\right)  - g^{R}_{biH}m_bm_G^2C_2\right], 
	\crn 	\overline{a}^{nGH}_r&= \sum_{i=1}^9 -\frac{g_{ZGH}  g^{L*}_{aiG}}{m_G^2}\times g^{R}_{biH}m_a\left(C_{00} +m_G^2C_1 \right), 
	\crn 	\overline{b}^{nGH}_l &=\sum_{i=1}^9 -\frac{g_{ZGH}  g^{L*}_{aiG}}{m_G^2} \left[ m_a \left(g^{R}_{biH} m_bX_2 - g^{L}_{biH} m_{n_i}X_{01}  \right)\right], 
	\crn	\overline{b}^{nGH}_r &=\sum_{i=1}^9 -\frac{g_{ZGH}  g^{L*}_{aiG}}{m_G^2} \left[ g^{R}_{biH} \left(m^2_{n_i}X_0 +m^2_a X_1 -2m_G^2 C_1\right) -g^{L}_{biH} m_{n_i}m_bX_2 \frac{}{}\right]. 
\end{align}
The contributions of the two diagrams {(3) and (4)} are suppressed {due to the} tiny mixing in $g_{ZWh_1}$ and {the} large $m_G=m_{Y}\gg m_W$. 

The form factors  relating to diagram {(4)} are: 
\begin{align}
	\label{eq:abFVH}
	\overline{a}^{nHG}_l &=\sum_{i=1}^9 -\frac{g_{ZGH}  g^{L}_{biG}}{m_G^2} \left[g^{L*}_{aiH} m_{n_i}\left( m_G^2 C_0 -C_{00}\right) - g_{aiH}^{R*} m_a m_G^2 C_1   \right], 
	\crn 	\overline{a}^{nHG}_r &= \sum_{i=1}^9 -\frac{g_{ZGH}  g^{L}_{biG}}{m_G^2} \left[   g_{aiH}^{R*} m_b\left(m_G^2 C_2 +C_{00}\right)\right], 
	\crn 	\overline{b}^{nHG}_l&= \sum_{i=1}^9 -\frac{g_{ZGH}  g^{L}_{biG}}{m_G^2} \left[ g^{R*}_{aiH}\left(m^2_{n_i} X_0 +m_b^2X_2- 2m_G^2C_2 \right) -g^{L*}_{aiH}m_am_{n_i}X_1\right], 
	\crn	\overline{b}^{nHG}_r &= \sum_{i=1}^9 -\frac{g_{ZGH}  g^{L}_{biG} m_b}{m_G^2} \left[ g^{R*}_{aiH}m_aX_1 - g^{L*}_{aiH}m_{n_i} X_{02} \right],
\end{align}
In this model, there are two classes of contributions, namely:
\begin{align}
	\label{eq:nWh1}
m_G=m_W,\; m_H=m_{h^\pm_1},\; g^L_{aiH}=-\frac{g \lambda^{L,1}_{ai}}{\sqrt{2}m_W},\; g_{ZGH}=g_{ZW^+h^-_1}= -\frac{s_{\theta} s_{2\beta} m_W}{t_W\sqrt{3-4 s_W^2}}\to 0,
\end{align}
and
\begin{align}
\label{eq:nYh2}
m_G=m_Y,\; m_H=m_{h^\pm_2},\; g^L_{aiH}=-\frac{g \lambda^{L,2}_{ai}}{\sqrt{2}m_W},\; g_{ZGH}=g_{ZY^+h^-_2}= -\frac{c_{\beta} c_{13} m_W}{c_Ws_W}.	
\end{align}

The scalar exchanges, determined based on previous works \cite{Jurciukonis:2021izn, Hong:2023rhg, Hong:2024yhk}, give one-loop contributions. 
Form factors corresponding to diagrams {(5)} are:
\begin{align}
	%
	\bar{a}^{nss}_{l}&=- \frac{g^2}{m_W^2} \sum_{k=1}^2\sum_{i=1}^9  g_{Zh^+_kh^-_k} \lambda^{L,k*}_{ai} \lambda^{L,k}_{bi} C_{00},
	\crn  \bar{a}^{nss}_{r}& =-  \frac{g^2}{m_W^2} \sum_{k=1}^2 \sum_{i=1}^9  g_{Zh^+_kh^-_k} \lambda^{R,k*}_{ai} \lambda^{R,k}_{bi} C_{00}, 	\label{eq_ab4LR}
	\\  \bar{b}^{nss}_{l}& =-  \frac{g^2}{m_W^2} \sum_{k=1}^2 \sum_{i=1}^9g_{Zh^+_kh^-_k}  \left[   m_a  \lambda^{L,k*}_{ai} \lambda^{L,k}_{bi} X_1  +  m_b  \lambda^{R,k*}_{ai} \lambda^{R,k}_{bi} X_2  -m_{n_i} \lambda^{R,k*}_{ai} \lambda^{L,k}_{bi} X_0  \right] ,
	\crn  \bar{b}^{nss}_{r}& = - \frac{g^2}{m_W^2} \sum_{k=1}^2 \sum_{i=1}^9g_{Zh^+_kh^-_k} \left[   m_a  \lambda^{R,k*}_{ai} \lambda^{R,k}_{bi} X_1  +  m_b  \lambda^{L,k*}_{ai} \lambda^{L,k}_{bi}X_2  -m_{n_i} \lambda^{L,k*}_{ai} \lambda^{R,k}_{bi} X_0  \right],\nn
\end{align}
where {$g_{Zh^+_kh^-_k}$ is} given in Table \ref{table_ZBB}, and $\lambda^{XY,k}_{ab}\equiv \lambda^{X,k*}_{ai} \lambda^{Y,k }_{bj}$ with $X,Y=L,R$. The arguments  for PV-functions are $(m_{n_i^2}, m_{h^\pm_k}^2, m_{h^\pm_k}^2)$. 

The contributions from  diagram (6) is 
\begin{align}
	\label{eq_ab6LR}	
	\bar{a}^{snn}_{l}=& -\frac{g^2}{2 m_W^2} \sum_{k=1}^2 \sum_{i,j=1}^9 \left\{  G_{ij}\left[ \lambda^{LL, k}_{ab}  m_{n_i} m_{n_j}C_0 + \lambda^{RL, k}_{ab}  m_{a} m_{n_j}(C_0+C_1) \frac{}{}
	\right.\right.\crn&\left.\left.\qquad \qquad \qquad
	+  \lambda^{LR, k}_{ab}  m_{b} m_{n_i}(C_0+C_2) +  \lambda^{RR,k}_{ab}  m_{a} m_bX_0 \frac{}{}\right]   
	\right.\crn  &\left. \qquad \quad \frac{}{}-G_{ji} \left[ -\lambda^{LL,k}_{ab} \left( 2C_{00} +m_a^2 X_1 +m_b^2X_2 -m_Z^2 C_{12}\right) \frac{}{}
	\right.\right.\crn&\left.\left.\qquad \qquad \qquad \; \frac{}{}-m_am_{n_i} \lambda^{RL,k}C_1 -m_bm_{n_j} \lambda^{LR,k}_{ab} C_2 \right]\right\}
	\crn  \bar{a}^{snn}_{r}=& -\frac{g^2}{2 m_W^2} \sum_{k=1}^2 \sum_{i,j=1}^9 \left\{  G_{ij} \left[ -\lambda^{RR,k}_{ab} \left( 2C_{00} +m_a^2 X_1 +m_b^2X_2 -m_Z^2 C_{12}\right) \frac{}{}
	\right.\right.\crn&\left.\left.\qquad \qquad \qquad\; -\lambda^{LR,k}_{ab} m_am_{n_i}C_1 -\lambda^{RL,k}m_bm_{n_j}C_2 \frac{}{}\right]
	\right.\crn  &\left. \qquad \quad 
	\frac{}{} -G_{ji} \left[ \lambda^{RR, k}_{ab}  m_{n_i} m_{n_j}C_0 + \lambda^{LR, k}_{ab}  m_{a} m_{n_j}(C_0+C_1) \frac{}{}
	\right.\right.\crn&\left.\left.\qquad \qquad \qquad \frac{}{}+  \lambda^{RL,k }_{ab}  m_{b} m_{n_i}(C_0+C_2) +  \lambda^{LL,k}_{ab}  m_{a} m_bX_0\right] 	\right\},
	\crn \bar{b}^{snn}_{l}=& -\frac{g^2}{ m_W^2} \sum_{k=1}^2 \sum_{i,j=1}^9  \left[\frac{}{}G_{ij}  \left( \lambda^{RL,k }_{ab} m_{n_j}C_2 +  \lambda^{RR,k }_{ab}  m_{b} X_2\right) 
	%
	-G_{ji}  \left(  \lambda^{RL,k }_{ab} m_{n_i}C_1 +  \lambda^{LL,k }_{ab}  m_{a} X_1\right) 
	\right]	,
	\crn\bar{b}^{snn}_{r} =&- \frac{g^2}{ m_W^2} \sum_{k=1}^2  \sum_{i,j=1}^9 \left[\frac{}{} G_{ij}   \left(  \lambda^{LR,k }_{ab} m_{n_i}C_1 +  \lambda^{R R,k}_{ab}  m_{a} X_1\right)
	%
	-G_{ji} \left( \lambda^{LR,k }_{ab} m_{n_j}C_2 +  \lambda^{LL,k }_{ab}  m_{b} X_2\right) 
	\right], 
\end{align}
where $\lambda^{XY,k}_{ab}\equiv \lambda^{X,k*}_{ai} \lambda^{Y,k }_{bj}$ with $X,Y=L,R$. The PV-functions and their linear combinations are $F=F(m_a^2,m_Z^2,m_b^2; m_{h^\pm_k}^2,m^2_{n_i},m^2_{n_j})$ for all $F=C_{00,0,i,ij},X_{0,i}$ with $i,j=1,2$.  

Sum of two diagrams (9) and (10) gives the following non-zero contributions 
\begin{align}
	\label{eq_a910LR}
	\bar{a}^{ns}_{l}&= - \frac{ g^2g^{L}_{Ze}}{2 m_W^2(m_a^2 -m_b^2)}\sum_{k=1}^2 \sum_{i=1}^9  \left[  m_{n_i} \left( m_a \lambda^{RL,k}_{ab}  + m_b \lambda^{LR,k}_{ab}   \right) \left(B^{(1)}_0 -B^{(2)}_0\right) 
	\right. \crn& \left.  \hspace{5.2cm}  -m_am_b \lambda^{RR,k}_{ab} \left( B^{(1)}_1-B^{(2)}_1\right)   - \lambda^{LL,k }_{ab}   \left(m_a^2B^{(1)}_1 -m_b^2 B^{(2)}_1 \right) 
	\right],
	\crn \bar{a}^{ns}_{r}&= -\frac{ g^2g^{R}_{Ze}}{2 m_W^2(m_a^2 -m_b^2)} \sum_{k=1}^2 \sum_{i=1}^9 \left[  m_{n_i} \left( m_a \lambda^{LR,k}_{ab}  + m_b \lambda^{RL,k }_{ab}   \right) \left(B^{(1)}_0 -B^{(2)}_0 \right) 
	\right. \crn& \left. \hspace{5.2cm} -m_am_b \lambda^{LL,k}_{ab} \left( B^{(1)}_1 -B^{(2)}_1\right)   - \lambda^{RR,k}_{ab}   \left( m_a^2B^{(1)}_1 -m_b^2 B^{(2)}_1\right) 
	\right],
\end{align}
where $\lambda^{XY,k}_{ab}=\lambda^{X,k*}_{ai} \lambda^{Y,k}_{bi} $ with $X,Y=L,R$. The PV-functions are $B^{(l)}_{0,1}=B_{0,1}(p_l^2;m_{n_i}^2,m_{h^\pm_k}^2)$ with $k,l=1,2$. 
In the limit $\theta=0$, we derived  that
\begin{align*}
&\mathrm{div}[\bar{a}^{nh^+_1h^+_1}_{l}]=A (m_Dm_D^{\dagger})_{ba} \times (-t_L), \; \mathrm{div}[\bar{a}^{h^+_1nn}_{l}]=0, \; \mathrm{div}[\bar{a}^{nh^+_2}_{l}]=A (m_Dm_D^{\dagger})_{ba} \times (t_L),
\\ &\mathrm{div}[\bar{a}^{nh^+_2h^+_2}_{l}]=A (m_Dm_D^{\dagger})_{ba} \times (-t_W), \; \mathrm{div}[\bar{a}^{h^+_1nn}_{l}]=\frac{A (m_Dm_D^{\dagger})_{ba} }{2s_Wc_W}, \; \mathrm{div}[\bar{a}^{nh^+_2}_{l}]=A (m_Dm_D^{\dagger})_{ba}  (t_L),
\end{align*} 
where $ \; t_L = \dfrac{s^2_W-c^2_W}{2s_Wc_W}$ satisfy the divergent cancellation for the total amplitudes of LFV$Z$ decays.

\end{document}